\documentclass[journal,10pt]{IEEEtran}
\usepackage{graphicx,amsmath,amssymb}
\usepackage{cite}
\usepackage{subfigure}
\usepackage{cite}
\usepackage{fancyhdr}
\usepackage{mdwmath}
\usepackage{mdwtab}
\usepackage{balance}
\usepackage{xcolor}
\usepackage{bm}
\usepackage{amsthm}
\usepackage{algorithm}
\usepackage{algorithmic}
\usepackage{multirow}
\usepackage{flafter}
\usepackage{epstopdf}
\usepackage[backref=false]{hyperref}

\newtheorem{theorem}{Theorem}

\newtheorem{lemma}{Lemma}

\newtheorem{corollary}{Corollary}

\newtheorem{assumption}{Assumption}

\begin{document}
\title{Secure Short-Packet Communications \\ for RIS-Assisted AAV Networks}
\author{Huiling Liu,~
        Junshan Luo,~
        Shilian Wang,~
        Fanggang Wang,~\IEEEmembership{Senior Member,~IEEE},\\
        Theodoros A. Tsiftsis,~\IEEEmembership{Senior Member,~IEEE},
        and Symeon Chatzinotas,~\IEEEmembership{Fellow,~IEEE}
\thanks{Part of this work has been submitted to IEEE ICASSP 2026.}
\thanks{H. Liu, J. Luo and S. Wang are with the College of Electronic Science and Technology, National University of Defense Technology, Changsha, 410073, China (email: liuhuiling23@nudt.edu.cn, ljsnudt@foxmail.com, wangsl@nudt.edu.cn).}
\thanks{F. Wang is with the School of Electronic and Information Engineering, Beijing Jiaotong University, Beijing, 100044, China (email: wangfg@bjtu.edu.cn).}
\thanks{T. A. Tsiftsis is with the Department of Electrical and Electronic Engineering, University of Nottingham Ningbo China, Ningbo 315100, China, and also with the Department of Informatics and Telecommunications, University of Thessaly, Lamia 35100, Greece (e-mail: tsiftsis@uth.gr).}
\thanks{S. Chatzinotas is with the Interdisciplinary Centre for Security, Reliability and Trust (SnT), University of Luxembourg, L-1855 Luxembourg, Luxembourg (email: symeon.chatzinotas@uni.lu).}
}
\maketitle
\begin{abstract}
Advancements toward 6G have intensified demands for ultra-reliable low-latency communication, positioning short-packet communications as a critical technology for autonomous aerial vehicle (AAV) networks. However, the open broadcast nature introduces significant security vulnerabilities. Although physical-layer security offers a low-complexity solution by exploiting wireless channel randomness, the AAV communication performance severely degrades in weak-coverage or non-line-of-sight scenarios. To overcome these limitations, this paper proposes a short-packet communications framework for AAV networks that leverages reconfigurable intelligent surfaces (RIS) with the aim of extending coverage and enhancing secrecy capabilities. Analytical frameworks are developed to evaluate the average secrecy throughput (AST) in finite blocklength constraints for both external and internal eavesdropping scenarios, which incorporates non-orthogonal multiple access with imperfect successive interference cancellation. Asymptotic approximations of AST are derived as transmit power approaches infinity. Furthermore, we formulate a blocklength optimization problem to maximize the AST, effectively resolving the trade-offs among delay, reliability, and secrecy. Extensive simulations validate the analytical frameworks, which reveal that large-scale RIS deployment significantly boosts AST, and the power allocation coefficient exhibits dual effects in the internal eavesdropping scenario. These observations provide useful insights for designing reliable and secure low-latency AAV communications systems. 
\end{abstract}
\section{Introduction}
\IEEEPARstart {W}{ith} the rapid evolution of global communication technologies, the upcoming sixth-generation (6G) mobile network is anticipated to not only retain the high-speed and high-capacity features of 5G, but also achieve a quantum leap in ultra-reliable and low-latency communication (URLLC), advancing toward hyper-reliable and low-latency communication \cite{marco2024}. URLLC is essential for supporting applications extremely sensitive to latency and reliability, such as industrial automation control, telemedicine surgery, and autonomous driving. These use cases require data transmission to be completed within very short time frames with exceptionally low packet error rates, imposing stringent demands on the low-latency and high-reliability performance of communication networks. Short-packet communication emerges as a key technology to fulfill these requirements, focusing on the efficient and rapid transmission of compact data packets \cite{Durisi2016}. Short-packet communications enables the delivery of critical information within constrained time intervals, thereby supporting real-time operation for a variety of latency-sensitive applications.
\subsection{Technical Literature Review}
In short-packet communications, the Shannon capacity in the infinite blocklength regime becomes inapplicable. A foundational breakthrough was made by Polyanskiy \textit{et al.} \cite{Polyanskiy2010}, which derived an analytical expression for the maximum achievable rate and reliability of data transmission in the finite blocklength regime. Subsequent studies have built upon this work. For instance, \cite{Fangchao2016} and \cite{Yulin2015} investigated the performance of short-packet communications networks in point-to-point and relay-assisted scenarios. In addition, closed-form expressions for the average block error rate (BLER) over Nakagami-$m$ fading channels were derived for both single-antenna and multi-antenna systems in \cite{Jianchao2019} and \cite{Tran2021}. The results in \cite{Polyanskiy2010} further facilitate the resource allocation in short-packet communications systems, enabling efficient utilization of limited blocklength and reducing transmission loss. The authors in \cite{jiechen2019} showed that improving the signal-to-interference-plus-noise ratio (SINR) improves the transmission rate and reduces BLER simultaneously. Such improvement can be achieved through multi-antenna spatial diversity techniques \cite{yifangu2018}. Furthermore, researches on short-packet communications have covered many dimensions, such as hybrid long-short-packet communications \cite{jiayao2024}, energy efficiency \cite{dingxu2024}, age of information \cite{haoxu2024} and integrated sensing and communication \cite{ning2024}, etc.

As space-based nodes, autonomous aerial vehicle (AAV) has significant potential in applications such as emergency communication, logistics and distribution, wide-area monitoring, etc., making them ideal candidates for URLLC deployments. In a typical AAV control scenario, the ground control station (GCS) transmits control signals in the form of short packets to the AAV. These transmissions demand extremely low latency and high reliability to ensure real-time responsiveness of the AAV, which is critical for flight safety and mission accuracy. Recent years have witnessed growing research interest in short-packet communications for AAV networks. For example, \cite{kezhi2021} derived closed-form expressions for the average BLER and achievable throughput in both free-space and three-dimensional channel models in AAV networks employing short-packet communications. Meanwhile, \cite{jinxie2023} treated the AAV as a hybrid access point and jointly optimized AAV position, transmit power, and transmission time to maximize energy efficiency of the system.

However, AAV communications face significant challenges. Their open wireless nature and broadcast characteristics make them highly vulnerable to eavesdropping threats. Conventional physical-layer security schemes suffer from significant performance degradation in weak coverage and non-line-of-sight (NLoS) scenarios, often failing to guarantee both reliability and security \cite{khan2022}. To address these coverage and security challenges, reconfigurable intelligent surface (RIS) \cite{junshan2021} have recently emerged as a promising solution for AAV networks, demonstrating unique advantages \cite{xintong}. First, it can effectively enhance signal strength in weak coverage and NLoS, thereby improving connectivity in blind spots. Moreover, its intelligent and controllable reflection capability allows for the dynamic optimization of signal transmission paths, which not only improves communication efficiency but also strengthens security. Furthermore, RIS facilitates energy-efficient communications due to its passive elements, resulting in extremely low energy consumption. Several studies have investigated the deployment of RIS on building facades to assist AAV communications. For instance, \cite{Agrawal2022} investigated multiple-RIS selection strategies for short-packet communications in AAV networks, proposing two practical approaches including partial channel state information (low overhead) and full channel state information (high performance) and deriving their selection probabilities. The authors in \cite{Ranjha2021} investigated scenarios where a ground transmitter transmits short-packet data to a ground device via AAV and RIS, minimizing the BLER through joint optimization of RIS beamforming, AAV positions, and blocklength. In contrast, \cite{hong2023} and \cite{laiwei2024} investigated the integration of RIS on board the AAV, achieving intelligent reflection of aerial signals. Specifically, Ren \textit{et al.} utilized RIS to improve the energy efficiency of wireless power transmission systems for AAV \cite{hong2023}. Lai \textit{et al.} analyzed the average BLER in an AAV-mounted RIS-assisted short-packet communications network by incorporating random location of AAV and the fading channels of ground users within a limited range \cite{laiwei2024}.

Regarding security, Diao \textit{et al.} explored three distinct RIS deployment strategies for secure AAV communication networks, demonstrating that RIS-assisted approaches yield substantial performance gains compared to non-RIS scenarios \cite{danyu2024}. The authors of \cite{adam2024} and \cite{wenjing2023} investigated the problem of secure multi-user communication for AAV carrying RIS. \cite{Majidtcom} analyzed the impact of varying the number of AAV and the location of eavesdroppers on secrecy transmission. Pang \textit{et al.} maximized the average secrecy rate by optimizing the trajectory of the AAV, the transmit beamforming, and the phase shift of the RIS \cite{xiaoweitcom}. In order to improve spectral efficiency, non-orthogonal multiple access (NOMA) has been introduced into the AAV networks. NOMA allows multiple users to share the same spectrum resources through power-domain or code-domain multiplexing, which significantly improves system capacity and spectral efficiency \cite{huilingtvt}. Its core mechanism lies in the superposition transmission at the transmitter and successive interference cancellation (SIC) at the receiver \cite{qunshutcom}. The work in \cite{globecom} verified the superiority of an RIS-assisted secure NOMA transmission scheme for AAV over conventional RIS-aided AAV networks. Furthermore, the work in \cite{jianiotj} jointly optimized AAV swarm trajectories, inter-AAV power allocation, and RIS reflection coefficients to maximize the overall security rate.
\subsection{Motivation}
However, the majority of existing studies on RIS-assisted AAV communications rely on infinite blocklength assumptions, which do not align with the low-latency requirements of short-packet communications. The security performance of RIS-assisted AAV networks in finite blocklength constraints remains underexplored, with limited analysis of the associated performance loss. Moreover, there is a lack of effective theoretical or design frameworks for systematically optimizing blocklength to balance secrecy, latency, and reliability in RIS-aided AAV communications. These research gaps motivate the present work. 

In this paper, we conduct a comprehensive investigation into the secrecy performance of RIS-assisted AAV short-packet communications networks. The station transmits short control signals to AAV via RIS in the presence of a ground eavesdropper. We develop analytical frameworks for the average secrecy throughput (AST) in both external and internal eavesdropping scenarios. Using the derived approximations, we further propose a blocklength optimization strategy to maximize the AST. 

The main contributions of this work are as follows:
\begin{itemize}
\item Departing from conventional infinite-blocklength models in RIS-AAV security research, this work introduces the first secure short-packet communications framework for RIS-assisted AAV networks. The model encompasses two critical eavesdropping threats: external eavesdroppers and internal adversaries, where the latter exploits NOMA with imperfect SIC, filling a gap in URLLC security analysis.
\item We establish comprehensive analytical frameworks to evaluate the secrecy performance of the proposed RIS-assisted AAV short-packet communications networks in finite blocklength constraints. Closed-form expressions for the AST are derived in both external and internal eavesdropping scenarios, solving the challenge of quantifying security-reliability-delay tradeoffs in practical short-packet regimes. These analytical results explicitly characterize the joint impact of blocklength and RIS deployment parameters on AST. To provide fundamental insights, asymptotic AST expressions at high transmit power are derived, while benchmarking against the infinite-blocklength regime reveals convergence behaviors and fundamental performance limits.
\item We propose blocklength optimization algorithms to maximize the AST, addressing both unconstrained scenarios and delay/reliability constrained regimes. For constrained cases, delay and reliability requirements are converted to mathematically tractable closed-form bounds on blocklength, enabling efficient one-dimensional optimization. This framework effectively balances the trade-offs among latency, reliability, and security.
\item Numerical results validate the theoretical analysis and quantitatively characterize critical parameters affecting the secrecy performance. It reveals that for both eavesdropping scenarios: 1) Blocklength and bits-per-block exhibit dual effects on AST, which initially enhances and then degrades beyond optimal thresholds. 2) Large-scale RIS deployment consistently improves AST. 3) Unique AST-maximizing blocklengths exist in both scenarios. Furthermore, the impact of transmission power diverges. It enhances AST for external eavesdropping but yields diminishing returns against internal threats due to SIC-induced error propagation.
\end{itemize}

\begin{figure}[!t]
\setlength{\abovecaptionskip}{0pt}
\centering
\includegraphics [width=3.7in]{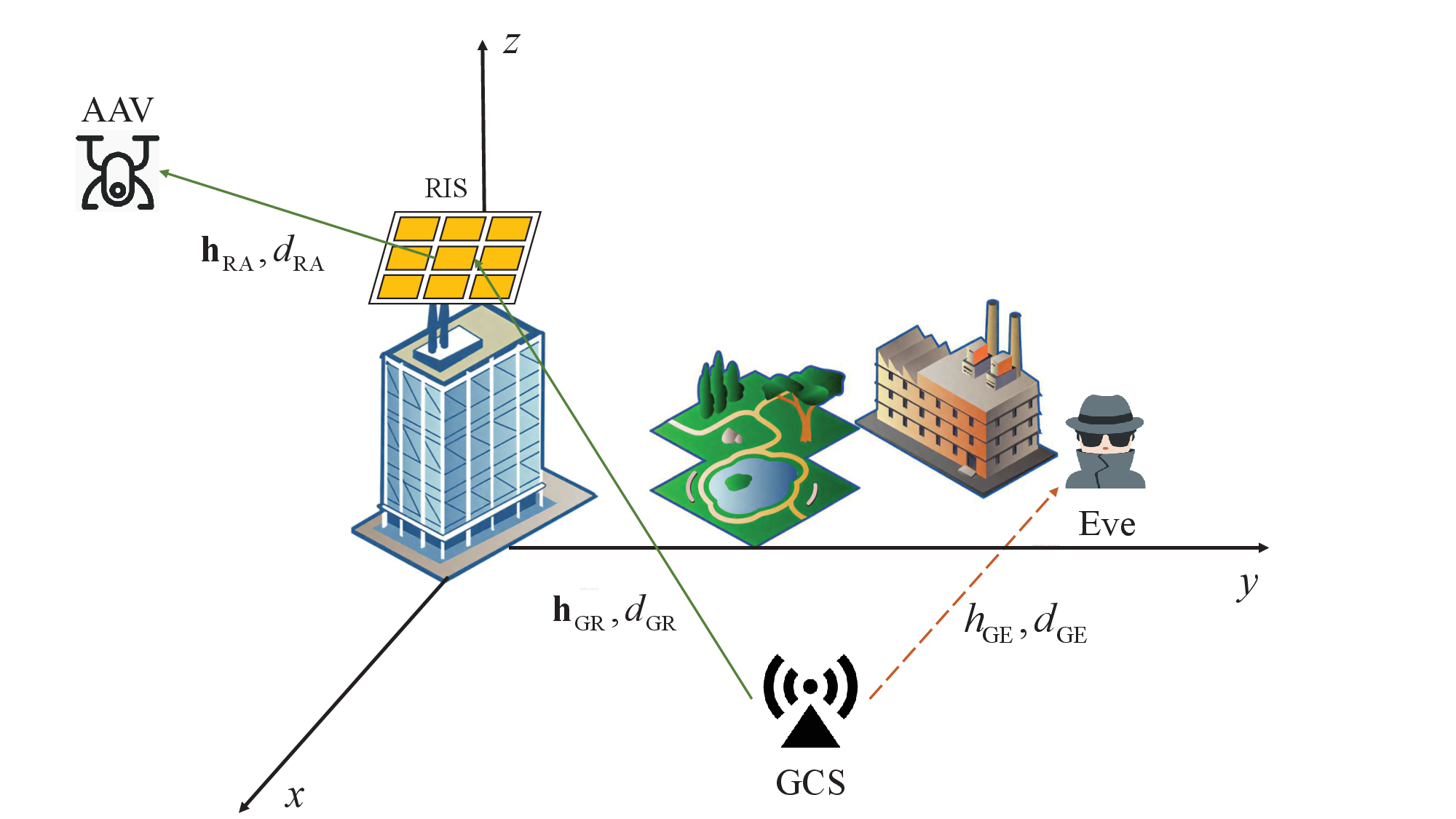}
\caption{System model of the RIS-assisted AAV wireless network with secure short-packet communication.}
\label{fig1}
\end{figure}

The remainder of the paper is summarized as follows. Section II introduces the system model and signal model. The performance metrics used in this work are   described in Section III. Section IV derives the AST achievable by AAV in two eavesdropping scenarios. In Section V, we optimize the blocklength to maximize the asymptotic expression of AST derived in Section IV. Section VI presents the simulations. Finally, the paper is concluded in Section VII.

\emph{Notation:} $\mathbb{C}^{N\times M}$ denotes an $N\times M$ space of complex matrices. $\mathbb{E} \left[  \cdot  \right]$ is expectation operation. ${f_X}\left( x \right)$ and ${F_X}\left( x \right)$ represent the probability density function (PDF) and cumulative distribution function (CDF) of a random variable $X$, respectively. diag$\left( \cdot \right)$ is a diagonal matrix. ${\cal C}{\cal N}\left( {a,b} \right)$ denotes a complex Gaussian distribution with mean $a$ and variance $b$. $\mathrm{Rician}\left( K, \Omega \right)$ is Rician fading distribution with $K$ as the Rician K-factor and $\Omega$ as the statistical average of the channel power gain. $\mathbb{P} \left\{  \cdot  \right\}$ denotes the probability calculation. $\left\lfloor x \right\rfloor $ and $\left\lceil x \right\rceil $ represent rounding down and up for $x$, respectively.
\section{System Model}
\subsection{Wireless Communication System Description}
Consider an RIS-assisted AAV secure short-packet communications network, as shown in Fig. 1. Due to the complex ground environment, the station transmits control signals to the AAV via the RIS deployed on a tall building, in the presence of an eavesdropper (Eve). Two common eavesdropping threats are considered: external eavesdropper and internal untrusted user. An external eavesdropper is an unlawful external third-party node attempting to intercept GCS-AAV communications, with the AAV directly decoding its own \cite{huiming2019twc}. To improve resource utilization efficiency, the NOMA scheme is used in internal eavesdropping scenarios. Therefore, AAV, being a high-priority user, is required to process and decode signals from Eve before decoding its own signals \cite{Xiazhitcom}. All nodes are single antenna configurations except for the RIS, which comprises $N$ passive reflecting elements. The reflection matrix is denoted as ${\bf{\Theta }}  = \textrm{diag}\left( {{\eta _1}{e^{j{\theta _1}}},...,{\eta _n}{e^{j{\theta _n}}},...,{\eta _N}{e^{j{\theta _N}}}} \right)$, where ${\eta _n} \in \left( {0,1} \right]$ and ${\theta _n} \in \left[ {0,2\pi } \right)$ denote amplitude and phase of the $n$-th element, respectively. The complex channel coefficients from GCS to RIS, from RIS to AAV and from GCS to Eve are denoted as ${{\bf{h}}_{{\rm{GR}}}}\in\mathbb{C}^{N\times1}$, ${{\bf{h}}_{{\rm{RA}}}} \in\mathbb{C}{^{N \times 1}}$ and ${h_{{\rm{GE}}}}$ respectively\footnote{In this work, we assumes the RIS is deployed to assist the legitimate link and its phase shifts are configured to maximize the signal strength at the AAV. The path from the RIS to Eve is treated as one of the Rayleigh-distributed multipath components within the channel ${h_{{\rm{GE}}}}$.}. In urban environments, due to the scattering of the obstacles, the terrestrial wireless link ${h_{{\rm{GE}}}}$ is characterized by Rayleigh fading, following ${h_{{\rm{GE}}}} \sim {\cal C}{\cal N}\left( {0,1} \right)$. The links via RIS $\mathbf{h}_{\rm{GR}}$ and $\mathbf{h}_{\rm{RA}}$ are modeled using Rician fading distributions $\left| h_{\mathrm{GR},i} \right| \sim \mathrm{Rician}\left( K_{\mathrm{GR}}, \Omega_{\mathrm{GR}} \right)$ and $\left| h_{\mathrm{RA},i} \right| \sim \mathrm{Rician}\left( K_{\mathrm{RA}}, \Omega_{\mathrm{RA}} \right)$.
\subsection{Signal Model}
In order to maximize the SNR of the AAV, the RIS considers coherent combination and optimal phase shift, i.e., it is set to ${\theta _i} = \angle \left( {h_{GR,i}^ * h_{RA,i}^ * } \right)$ for $i = 1,...,n,...,N$, where ${h_{\mathrm{GR},i}}$ and ${h_{\mathrm{RA},i}}$ are the channels from the $i$-th RIS element to GCS and AAV, respectively. Furthermore, the reflection coefficient of the RIS element is set as ${\eta _n}=1$. Thus, the cascading channel through the RIS can be re-expressed as
\begin{align}
\label{eq1}
{\bf{h}}_{{\rm{RA}}}^{\rm{H}}{\bf{\Theta }}{{\bf{h}}_{{\rm{GR}}}} = \sum\limits_{i = 1}^N {\left| {{h_{{\rm{GR}},i}}{h_{{\rm{RA}},i}}} \right|} .
\end{align}

\textit{1) External Eavesdropping Scenario:} The signal-to-noise ratio (SNR) for AAV and Eve can be given, respectively, by
\begin{align}
\tilde \gamma _{\rm{A}}^{{\rm{Ex}}} &= {\rho _{\rm{A}}}d_{{\rm{RA}}}^{ - \alpha }d_{{\rm{GR}}}^{ - \alpha }{\bigg( {\sum\limits_{i = 1}^N {\left| {{h_{{\rm{GR}},i}}{h_{{\rm{RA}},i}}} \right|} } \bigg)^2} \label{eq2} \\ 
\tilde \gamma _{\rm{E}}^{{\rm{Ex}}} &= {\rho _{\rm{E}}}d_{{\rm{GE}}}^{ - \alpha }{\left| {{h_{{\rm{GE}}}}} \right|^2} 
\label{eq3}
\end{align}
where ${\rho _{\rm{A}}} = {P_{\rm{G}}}\sigma _{\rm{A}}^{ - 2}$ and ${\rho _{\rm{E}}} = {P_{\rm{G}}}\sigma _{\rm{E}}^{ - 2}$ denote the transmit SNR to the AAV and Eve, respectively. $P_{\rm{G}}$ is the transmit power of the station. $\sigma _{\rm{A}}^2$ and $\sigma _{\rm{E}}^2$ are the power of the additive white Gaussian noise at AAV and Eve, respectively. ${d_{\rm{RA}}}$, ${d_{\rm{GR}}}$ and ${d_{\rm{GE}}}$ represent the distances between RIS-AAV, GCS-RIS and GCS-Eve, respectively. $\alpha $ denotes the pathloss factor.

\textit{2) Internal Eavesdropping Scenario:} The station considers the use of NOMA scheme to improve the resource utilization efficiency, where the internal untrusted user is marked as Eve. The superimposed signal transmitted by station is given by
\begin{align}
x = \sqrt {{a_{\rm{A}}}} {x_{\rm{A}}} + \sqrt {{a_{\rm{E}}}} {x_{\rm{E}}} \label{eq4}
\end{align}
where ${a_{\rm{A}}}$ and ${a_{\rm{E}}}$ is the power allocation coefficient for AAV and Eve, respectively, with ${a_{\rm{A}}} + {a_{\rm{E}}} = 1$ and ${a_{\rm{A}}} < {a_{\rm{E}}}$.

From the decoding principle in the NOMA scheme, the AAV first decodes the signal of Eve, performs SIC, and then decodes its own signal. Consider the imperfect SIC, the first decoding step is prone to errors, implying that the signal interference of Eve cannot be completely eliminated. Consequently, the performance of the AAV is impaired, a phenomenon particularly prevalent in short-packet communications schemes. Thus, the SINR of decoding the signal of Eve and its own signal are characterized by
\begin{align}
\tilde \gamma _{{\rm{A}} \to {\rm{E}}}^{{\rm{In}}} &= \frac{{{a_{\rm{E}}}{\rho _{\rm{A}}}d_{{\rm{RA}}}^{ - \alpha }d_{{\rm{GR}}}^{ - \alpha }{\bigg( {\sum\limits_{i = 1}^N {\left| {{h_{{\rm{GR}},i}}{h_{{\rm{RA}},i}}} \right|} } \bigg)^2}}}{{{a_{\rm{A}}}{\rho _{\rm{A}}}d_{{\rm{RA}}}^{ - \alpha }d_{{\rm{GR}}}^{ - \alpha }{\bigg( {\sum\limits_{i = 1}^N {\left| {{h_{{\rm{GR}},i}}{h_{{\rm{RA}},i}}} \right|} } \bigg)^2} + 1}} \label{eq5} \\
\tilde \gamma _{\rm{A}}^{{\rm{In}}} &= \frac{{{a_{\rm{A}}}{\rho _{\rm{A}}}d_{{\rm{RA}}}^{ - \alpha }d_{{\rm{GR}}}^{ - \alpha }{\bigg( {\sum\limits_{i = 1}^N {\left| {{h_{{\rm{GR}},i}}{h_{{\rm{RA}},i}}} \right|} } \bigg)^2}}}{{\omega {{\left| {{h_{\rm{I}}}} \right|}^2}{\rho _{\rm{A}}} + 1}} \label{eq6}
\end{align}
respectively, where $\omega  \in \left[ {0,1} \right]$ represents the residual interference level for SIC. ${\omega}=0$ indicates that the perfect SIC is performed. ${{h_{\rm{I}}}}$ is the corresponding channel residual interference coefficient, following ${{h_{\rm{I}}}} \sim {\cal C}{\cal N}\left( {0,{\Omega _{\rm{I}}}} \right)$ \cite{Gong2022TVT}.

For the internal untrustworthy user, we assume the worst-case scenario where he can perfectly decode the first-stage signal $x_{\rm{E}}$. This conservative assumption holds in practice, particularly in high-SNR regime and when the power allocation factors of Eve $a_{\rm{E}}$ is sufficiently large \cite{Xiazhitcom}. Correspondingly, the SNR of Eve eavesdropping ${x_{\rm{A}}}$ is denoted as  
\begin{align}
\tilde \gamma _{{\rm{E}} \to {\rm{A}}}^{{\rm{In}}} = {a_{\rm{A}}}{\rho _{\rm{E}}}d_{{\rm{GE}}}^{ - \alpha }{\left| {{h_{{\rm{GE}}}}} \right|^2}. \label{eq7}
\end{align}

If the Eve treats $x_{\rm{E}}$ as interference and decodes $x_{\rm{A}}$ directly, it will result in an overestimation of the system secrecy performance. This is because the power allocated to Eve by the station is generally higher than AAV, i.e., ${a_{\rm{E}}} > {a_{\rm{A}}}$.

\section{Secure Transmission Metrics In Finite Blocklength Constraints}
In short-packet communication scenarios, the use of finite blocklength codes introduces fundamental limitations that significantly impact system performance, particularly in terms of latency, transmission reliability, and communication secrecy. To accurately capture and analyze these effects, this section introduces the secure performance metric known as AST, which is adopted throughout this work. We further elaborate on how the secrecy capacity is affected in the finite blocklength constraint, providing theoretical insights into the security-performance tradeoffs inherent in such regimes.

The achievable AST, measured in bits per channel use (BPCU), is defined as the reliable decoding rate at the legitimate user while ensuring secrecy against eavesdroppers. The expression is given by
\begin{align}
\tau  = \frac{B}{m}\left( {1 - \bar \varepsilon } \right) \label{eq8}
\end{align}
where $B$ denotes the number of bits transmitted per block; $m$ is the blocklength; $\varepsilon $ is the BLER with secrecy requirement and $\bar \varepsilon  = {\mathbb{E}}\left[ \varepsilon  \right]$ denotes its average value.

Distinct from the conventional secrecy throughput in the infinite blocklength regime \cite{Gong2022TVT}, the AST in the finite blocklength regime captures the fundamental tradeoffs among reliability, security, and latency. A critical step in characterizing the AST is the derivation of the BLER with secrecy requirement. We outline this derivation below.

First, we review the BLER in finite blocklength conditions which is denoted by $\varepsilon$. Based on finite blocklength channel coding theory \cite{Polyanskiy2010}, for given blocklength $m$ and BLER $\varepsilon $, the maximum achievable rate is approximated by
\begin{align}
R\left( {m,\varepsilon } \right) = C - \sqrt {\frac{V}{m}} {Q^{ - 1}}\left( \varepsilon  \right) + \mathcal{O}\left( {\frac{{\log m}}{m}} \right) \label{eq9}
\end{align}
where $C = {\log }\left( {1 + \tilde \gamma } \right)$ denotes the Shannon capacity and $\tilde \gamma$ is the SNR; ${Q^{ - 1}}\left( x \right)$ is the inverse of Gaussian Q-function; $V = \big( {1 + {{\left( {1 + \tilde \gamma } \right)}^{ - 2}}} \big){\log ^2}e$ is channel dispersion and $\mathcal{O}\left(  \cdot  \right)$ is the higher order terms. When $m \to \infty $, $R$ approximates the Shannon capacity $C$ because the modification terms due to the finite blocklength can be ignored.

From \eqref{eq9}, the BLER can be expressed as
\begin{align}
\varepsilon  = Q\left( {\sqrt {\frac{V}{m}} \left( {\log \left( {1 + \tilde \gamma } \right) - \frac{B}{m}} \right)} \right)
\end{align}

From wiretap coding theorem of Wyner, both decoding error and information leakage asymptotically approach zero as the blocklength approaches infinity, provided the legitimate channel is stronger than that of the eavesdropper. However, in the finite blocklength regime, decoding errors and information leakage cannot be ignored.

The BLER with secrecy requirement can be derived from the finite blocklength secrecy rate. Specifically, as shown in \cite{weiyang2016}, on the maximum achievable secrecy rate for a given block length $m$ in the constraints of decoding error probability $\varepsilon $ and information leakage probability $\delta $ is approximated by
\begin{align}
{R^{\rm{s}}}\left( {m,\varepsilon ,\delta } \right) = {C_{\rm{s}}} - \sqrt {\frac{{{V_{\rm{A}}}}}{m}} {Q^{ - 1}}\left( \varepsilon  \right) - \sqrt {\frac{{{V_{\rm{E}}}}}{m}} {Q^{ - 1}}\left( \delta  \right) \label{eq11}
\end{align}
conditional on ${\tilde \gamma _{\rm{A}}} > {\tilde \gamma _{\rm{E}}}$; otherwise ${R^{\rm{s}}}=0$. In \eqref{eq11}, ${C_{\rm{s}}} = {\log }\left( {1 + {\tilde \gamma _{\rm{A}}}} \right) - {\log }\left( {1 + {\tilde \gamma _{\rm{E}}}} \right)$ denotes the infinite-blocklength secrecy rate. ${V_{\rm{A}}} = \big( {1 - {{\left( {1 + {{\tilde \gamma }_{\rm{A}}}} \right)}^{ - 2}}} \big){\log ^2}e$ and ${V_{\rm{E}}} = \big( {1 - {{\left( {1 + {{\tilde \gamma }_{\rm{E}}}} \right)}^{ - 2}}} \big){\log ^2}e$ are the channel dispersion of the legitimate and eavesdropping channels, respectively.  The second and third terms quantify the penalty in reliability and security due to finite blocklength.

From \eqref{eq11}, the instantaneous BLER with secrecy requirement of the legitimate communication, constrained by the information leakage probability $\delta$ is given by \cite{huiming2019twc}
\begin{align}
\varepsilon  = Q\left( {\sqrt {\frac{m}{{{V_A}}}} \left( {\log \frac{{1 + {{\tilde \gamma }_{\rm{A}}}}}{{1 + {{\tilde \gamma }_E}}} - \sqrt {\frac{{{V_{\rm{E}}}}}{m}} {Q^{ - 1}}\left( \delta  \right) - \frac{B}{m}} \right)} \right) \label{eq12}
\end{align}
for ${\tilde \gamma _{\rm{A}}} > {\tilde \gamma _{\rm{E}}}$; otherwise $\varepsilon=1$.

We fix the information leakage probability ${\delta}$ in $\left( {0,\frac{1}{2}} \right)$ to impose a strict secrecy constraint. If $\delta  > \frac{1}{2}$, the last term in \eqref{eq11} is negative. Both the number of bits transmitted per block $B$ and the blocklength $m$ directly affect the physical-layer transmission delay in channel uses. Therefore $B$ and $m$ should be chosen carefully. In the subsequent sections, we analyze the effects of $B$ and $m$ on AST. Moreover, for a fixed $B$, AST can be maximized by optimally selecting $m$.

\section{Closed-Form Characterization and Asymptotic Behavior of Secrecy Performance}
This section derives closed-form analytical expressions for the AST in both internal and external eavesdropping scenarios, based on the channel statistical characteristics. To provide more insights, we further analyze the asymptotic behavior of the AST in the high-SNR regime and compare the performance with the infinite blocklength case.

\subsection{Channel Statistical Characteristics}
Let the cascade channel be $X = \sum\nolimits_{i = 1}^N {\left| {{h_{{\rm{GR}},i}}{h_{{\rm{RA}},i}}} \right|} $ and apply the series of the Laguerre polynomial to approximate its PDF and CDF as 
\begin{align}
{f_X}\left( x \right) &= {\Gamma ^{ - 1}}\big( {\hat k} \big){\hat \theta ^{ - \hat k}}{x^{\hat k - 1}}{e^{ - \frac{x}{{\hat \theta }}}} \label{eq13} \\
{F_X}\left( x \right) &= {\Gamma ^{ - 1}}\big( {\hat k} \big)\gamma \big(\hat k,{{\hat \theta }^{ - 1}}x \big) \label{eq14}
\end{align}
where $\hat k = \frac{{{\mu ^2}N}}{\vartheta }$; $\hat \theta  = \frac{\vartheta }{\mu }$; $\mu  = \frac{{\pi \sqrt {{\Omega _{{\rm{RA}}}}{\Omega _{{\rm{GR}}}}} }}{{4\left( {\kappa  + 1} \right)}}L_{\frac{1}{2}}^2\left( { - \kappa } \right)$; $\vartheta  = {\Omega _{{\rm{RA}}}}{\Omega _{{\rm{GR}}}}\big( {1 - \frac{{{\pi ^2}}}{{16{{\left( {\kappa  + 1} \right)}^2}}}L_{\frac{1}{2}}^4\left( { - \kappa } \right)} \big)$; $\kappa  = {K_{\rm{GR}}} = {K_{\rm{RA}}}$; $L_{\frac{1}{2}}\left(  \cdot  \right)$ denotes the Laguerre polynomial; $\gamma \left( {a,x} \right)$ is the lower incomplete Gamma function and $\Gamma \left(  \cdot  \right)$ is the gamma function. There are parallel to the Lemma 2 in \cite{zhenwen2024tcom}.

Let $Z = {\big( {\sum\nolimits_{i = 1}^N {\left| {{h_{{\rm{GR}},i}}{h_{{\rm{RA}},i}}} \right|} } \big)^2}$ and when $X > 0$, then the CDF of $Z$ can be derived as
\begin{align}
{F_Z}\left( z \right) = {\Gamma ^{ - 1}}\big( {\hat k} \big)\gamma \big( {\hat k,{{\hat \theta }^{ - 1}}\sqrt z } \big). \label{eq15}
\end{align}

Based on the statistical characterization of the cascade channel, we calculate the statistical characterizations of $\tilde \gamma _{\rm{A}}^{{\rm{Ex}}}$, $\tilde \gamma _{{\rm{A}} \to {\rm{E}}}^{{\rm{In}}}$ and $\tilde \gamma _{\rm{A}}^{{\rm{In}}}$, respectively in the following.

\textit{1) Statistical Characteristic of $\tilde \gamma _{\rm{A}}^{{\rm{Ex}}}$}: Its CDF is obtained by \eqref{eq15} and some mathematical operations, and is given by
\begin{align}
{F_{{\tilde \gamma _{\rm{A}}^{{\rm{Ex}}}}}}\left( {\bar x } \right) = {\Gamma ^{ - 1}}\big( {\hat k} \big)\gamma \big( {\hat k,{{\hat \theta }^{ - 1}}\sqrt {{\Delta ^{ - 1}}\bar x} } \big) \label{eq16}
\end{align}
where $\Delta  = {\rho _{\rm{A}}}d_{{\rm{RA}}}^{ - \alpha }d_{{\rm{GR}}}^{ - \alpha }$.

\textit{2) Statistical Characteristic of $\tilde \gamma _{{\rm{A}} \to {\rm{E}}}^{{\rm{In}}}$}: Assuming $\bar Z = \frac{{{a_{\rm{E}}}\Delta Z}}{{{a_{\rm{A}}}\Delta Z + 1}}$, the CDF of $\tilde \gamma _{{\rm{A}} \to {\rm{E}}}^{{\rm{In}}}$ is obtained by
\begin{align}
{F_{\tilde \gamma _{{\rm{A}} \to {\rm{E}}}^{{\rm{In}}}}}\left( \bar z \right) = \mathbb{P} \bigg\{ {\bar Z < \frac{{\bar z}}{{{a_{\rm{E}}}\Delta  - {a_{\rm{A}}}\Delta \bar z}}} \bigg\}. \label{eq17}
\end{align}

By exploiting \eqref{eq15}, we obtain
\begin{align}\nonumber
{F_{\tilde \gamma _{{\rm{A}} \to {\rm{E}}}^{{\rm{In}}}}}\left( \bar z \right) &= {F_{\bar Z}}\bigg( {\frac{\bar z}{{{a_{\rm{E}}}\Delta  - {a_{\rm{A}}}\Delta \bar z}}} \bigg)\\
&= {\Gamma ^{ - 1}}\big( {\hat k} \big)\gamma \left( {\hat k,{{\hat \theta }^{ - 1}}\sqrt {\frac{{\bar z}}{{{a_{\rm{E}}}\Delta  - {a_{\rm{A}}}\Delta \bar z}}} } \right) \label{eq18}
\end{align}
which should be satisfied with ${a_{\rm{E}}}\Delta  - {a_{\rm{A}}}\Delta \bar z > 0$.

\textit{3) Statistical Characteristic of $\tilde \gamma _{\rm{A}}^{{\rm{In}}}$}: The CDF of $\tilde \gamma _{\rm{A}}^{{\rm{In}}}$ is given by the following lemma.

\emph{Lemma 1:} The CDF of $\tilde \gamma _{\rm{A}}^{{\rm{In}}}$ is given by
\begin{align}\nonumber
{F_{\tilde Z}}\left( {\tilde z} \right) \approx &\frac{{\pi {M_1}}}{{2\Gamma \big( {\hat k} \big)\omega {\rho _{\rm{A}}}{M_2}}} \\
&\quad \times \sum\limits_{i = 0}^{{M_2}} {\gamma \bigg( {\hat k,{{\hat \theta }^{ - 1}}\sqrt {\frac{{\tilde z\left( {{\zeta _i} + 1} \right)}}{{{A_1}}}} } \bigg)} \frac{{\sqrt {1 - {\varpi _i}^2} }}{{{\rm{ }}{e^{\frac{{{\zeta _i}}}{{\omega {\rho _{\rm{A}}}}}}}}}  \label{eq19}
\end{align}
where ${\zeta _i} = \frac{{{M_1}\left( {{\varpi _i} + 1} \right)}}{2}$, ${\varpi _i} = \cos {\frac{{\left( {2i - 1} \right)\pi }}{{2{M_2}}}}$ and ${M_2}$ is a complexity-vs-accuracy tradeoff parameter.

\emph{Proof: }The proof is provided in Appendix \ref{app:lemma1}. \hfill $\blacksquare$      

In addition, the PDF of the SNR for Eve in the external eavesdropping scenario is denoted as
\begin{align}
{f_{{\tilde \gamma _{\rm{E}}^{{\rm{Ex}}}}}}\left( {\bar y} \right) = {\left( {{\rho _{\rm{E}}}d_{{\rm{GE}}}^{ - \alpha }} \right)^{ - 1}}{{\mathop{ e}\nolimits} ^{ - \frac{{\bar y}}{{{\rho _{\rm{E}}}d_{{\rm{GE}}}^{ - \alpha }}}}}. \label{eq20}
\end{align}

Similarly, the PDF of the SNR for Eve in the internal eavesdropping scenario is given by
\begin{align}
{f_{\tilde \gamma _{{\rm{E}} \to {\rm{A}}}^{{\rm{In}}}}}\left( {\hat y} \right) = {\left( {{a_{\rm{A}}}{\rho _{\rm{E}}}d_{{\rm{GE}}}^{ - \alpha }} \right)^{ - 1}}{\mathop{ e}\nolimits} ^{- \frac{{\hat y}}{{{a_{\rm{A}}}{\rho _{\rm{E}}}d_{{\rm{GE}}}^{ - \alpha }}}}. \label{eq21}
\end{align}
\subsection{AST of AAV in External Eavesdropping Scenario}
From \eqref{eq8} and \eqref{eq12}, the AST of the AAV in external eavesdropping scenario is
\begin{align}
\tau ^{\rm{Ex}} = \frac{B}{m}\int_0^\infty  {\Psi \left( \bar y \right){f_{{\tilde \gamma _{\rm{E}}^{{\rm{Ex}}}}}}\left( \bar y \right)dy} \label{eq22}
\end{align}
where
\begin{align}
\Psi \left( \bar y \right) = \int_{\bar y}^\infty  {\big( {1 - {\varepsilon _{{\tilde \gamma _{\rm{A}}^{{\rm{Ex}}}} = \bar x|{\tilde \gamma _{\rm{E}}^{{\rm{Ex}}}}}}\left(\bar x \right)} \big){f_{{\tilde \gamma _{\rm{A}}^{{\rm{Ex}}}}}}\left(\bar x \right)d \bar x}. \label{eq23}
\end{align}

Note that the integral in \eqref{eq23} involves a Q-function, rendering the derivation highly challenging. Using the first-order Taylor approximation of the Q-function at $Q\left(\bar x \right) = \frac{1}{2}$, ${\varepsilon _{{\tilde \gamma _{\rm{A}}^{{\rm{Ex}}}} =\bar x|{\tilde \gamma _{\rm{E}}^{{\rm{Ex}}}}}}\left(\bar x \right)$ can be approximated as \eqref{eq24} at the top of the next page. In \eqref{eq24},\\
\begin{figure*}[!t]
\begin{align}
\label{eq24}
{\varepsilon _{{\tilde \gamma _{\rm{A}}^{{\rm{Ex}}}} =\bar x|{\tilde \gamma _{\rm{E}}^{{\rm{Ex}}}}}}\left(\bar x \right) \approx P\left(\bar x \right)= \left\{ \!\!\!\! {\begin{array}{*{20}{c}}
\begin{array}{l}
1,\\
\frac{1}{2} + k\left( {\bar x - {x_0}} \right),\\
0,
\end{array}&\begin{array}{l}
\bar x < \frac{1}{{2k}} + {x_0}\\
\bar x \in \left[ {\frac{1}{{2k}} + {x_0}, - \frac{1}{{2k}} + {x_0}} \right]\\
\bar x >  - \frac{1}{{2k}} + {x_0}
\end{array}
\end{array}} \right..
\end{align}
\hrulefill \vspace*{0pt}
\end{figure*}
\begin{align}
{x_0} = {2^{\sqrt {\frac{{{V_{\rm{E}}}}}{m}} {Q^{ - 1}}\left( \delta  \right) + \frac{B}{m}}}\left( {1 + \tilde \gamma _{\rm{E}}^{{\rm{Ex}}}} \right) - 1 \label{eq25}
\end{align}
\begin{align}\nonumber
k & = \left.\frac{{\partial {\varepsilon _{\tilde \gamma _{\rm{A}}^{{\rm{Ex}}} = \bar x|\tilde \gamma _{\rm{E}}^{{\rm{Ex}}}}}\left( {\bar x} \right)}}{{\partial \bar x}} \right|_{\bar x = {x_0}} \\
&=  - \sqrt {\frac{m}{{2\pi {x_0}\left( {{x_0} + 2} \right)}}}. \label{eq26}
\end{align}

Substituting \eqref{eq24} - \eqref{eq26} into \eqref{eq23}, $\Psi \left( \bar y \right)$ is expressed as $\Psi \left( \bar y \right) \approx \int_{\bar y}^\infty  {\left( {1 - P\left(\bar x \right)} \right){f_{{\tilde \gamma _{\rm{A}}^{{\rm{Ex}}}}}}\left(\bar x \right)d \bar x} $. Notice that the challenge now stems from the integral $\Psi \left( \bar y \right)$ having a non-constant lower limit $\bar y$. The following lemma justifies the approximation that shifts the lower integration limit from $\bar y$ to $0$.

\emph{Lemma 2:} $\Psi \left( \bar y \right)$ in \eqref{eq23} can be approximated as
\begin{align}
\Psi \left( \bar y \right) \approx \int_0^\infty  {\left( {1 - {\varepsilon _{{\tilde \gamma _{\rm{A}}^{{\rm{Ex}}}} = \bar x|{\tilde \gamma _{\rm{E}}^{{\rm{Ex}}}}}}\left(\bar x \right)} \right){f_{{\tilde \gamma _{\rm{A}}^{{\rm{Ex}}}}}}\left(\bar x \right)d \bar x}. \label{eq27}
\end{align}

\emph{Proof: }The proof is provided in Appendix B.  \hfill $\blacksquare$

Leveraging the above discussion, Theorem 1 gives an approximate expression for AST as follows.

\emph{Theorem 1:} In the external eavesdropping scenario, the AST that the AAV is able to achieve is given by \eqref{eq28} at the top of this page.
\begin{figure*}[!t]
\begin{align}
{\tau ^{{\rm{Ex}}}} \approx \frac{B}{m}\left( {1 - \Lambda \sum\limits_{i = 0}^{{M_2}} {\gamma \left( {\hat k,{{\left( {\frac{{\Xi  - 1 + \Xi {\zeta _i}}}{{\Delta {{\hat \theta }^2}}}} \right)}^{\frac{1}{2}}}} \right){e^{ - \frac{{{\zeta _i}}}{{{\rho _{\rm{E}}}d_{{\rm{GE}}}^{ - \alpha }}}}}\sqrt {1 - {\varpi _i}^2} } } \right). \label{eq28}
\end{align}
\hrulefill \vspace*{0pt}
\end{figure*}
In \eqref{eq28}, $\Lambda  = \frac{{\pi {M_1}d_{{\rm{GE}}}^\alpha }}{{2\Gamma (\hat k){\rho _{\rm{E}}}{M_2}}}$ and $\Xi  = {2^{{m^{\frac{1}{2}}}{Q^{ - 1}}\left( \delta  \right) + \frac{B}{m}}}$.

\emph{Proof: }The proof is provided in Appendix C.  \hfill $\blacksquare$ 

\textbf{\emph{Remark 1:}} The blocklength $m$ plays a crucial role in determining the AST. It has a two-fold influence on the performance of AAV: increasing blocklength $m$ prolongs the delay but also reduces the BLER of AAV. Thus, the blocklength needs to be carefully and rationally chosen to realize the trade-off between secrecy, delay and reliability.

The AST result presented in Theorem 1 for AAV in the face of external eavesdropping holds for a blocklength $m$ of any length, but the expression is rather intricate. In the Corollary 1, we present the asymptotic behaviors in the high-SNR regime.

\emph{Corollary 1:}
When the transmit power $P_{\rm{G}}$ is sufficiently large, the AST in the high-SNR regime is given by
\begin{align}
{\tau ^{{\rm{Ex,asy}}}} \approx \frac{B}{m}\Bigg( {1 \!-\! {{\left( {\frac{{2{e^{{\Lambda _2}}}}}{{{\Lambda _1}{{\hat \theta }^2}}}} \right)}^{\frac{{\hat k}}{2}}}{e^{\frac{{{e^{{\Lambda _2}}}}}{{8{{\hat \theta }^2}}}}}{D_{ - \hat k}}\left( {\sqrt {\frac{{{e^{{\Lambda _2}}}}}{{2{\Lambda _1}{{\hat \theta }^2}}}} } \right)} \Bigg) \label{eq29}
\end{align}
where ${\Lambda _1} = \frac{{d_{{\rm{GE}}}^\alpha \sigma _{\rm{E}}^{\rm{2}}}}{{d_{{\rm{RA}}}^\alpha d_{{\rm{GR}}}^\alpha \sigma _{\rm{A}}^{\rm{2}}}}$, ${\Lambda _2} = {{m^{ - \frac{1}{2}}}{Q^{ - 1}}\left( \delta  \right)} + \frac{B}{m}$ and ${D_n}\left( z \right)$ is the parabolic cylinder function \cite[(9.240)]{Table}.  

\emph{Proof: }The proof is provided in Appendix D.  \hfill $\blacksquare$  

\textbf{\emph{Remark 2:}} The result reveals that in the high-SNR regime, the AST becomes independent of the transmit power and is governed by coding efficiency, system configurations, and channel conditions. This implies that simply increasing transmit power is ineffective for improving AST. Instead, optimizing spectral efficiency, expanding antenna arrays, or refining coding schemes is necessary.

To gain more insights into the asymptotic system performance and to illuminate the relationship between finite and infinite blocklength regimes, the following corollary focuses on the limiting case where the blocklength $m$ approaches to infinity. In the ideal case of infinite blocklength, the system reliably transmits information with zero error probability and zero information leakage probability as long as ${\tilde \gamma _{\rm{A}}^{{\rm{Ex}}}} > {\tilde \gamma _{\rm{E}}^{{\rm{Ex}}}}$. Secure throughput is the product of transmission rate $\frac{B}{m}$ and ${\tilde \gamma _{\rm{A}}^{{\rm{Ex}}}} > {\tilde \gamma _{\rm{E}}^{{\rm{Ex}}}}$ probability.

\emph{Corollary 2:} When $m \!\to\! \infty$, the secrecy throughput is characterized by the following
\begin{align}\nonumber
\tau _{m \to \infty }^{{\rm{Ex}}}
    & \!=\! \frac{B}{m}\mathbb{P}\left\{ {\tilde \gamma _{\rm{A}}^{{\rm{Ex}}}} > {\tilde \gamma _{\rm{E}}^{{\rm{Ex}}}} \right\} \\
    & \approx \frac{B}{m}\!\!\left( {\!1 \!-\! \Lambda \! \sum\limits_{i = 0}^{{M_2}} {\gamma \!\left( {\!\hat k,\sqrt {\frac{{{\zeta _i}}}{{\Delta {{\hat \theta }^2}}}} } \right)\frac{{\sqrt {1 \!- \varpi _i^2} }}{{{e^{\frac{{{\zeta _i}}}{{{\rho _{\rm{E}}}d_{{\rm{GE}}}^{ - \alpha }}}}}}}} } \right). \label{eq30}
\end{align}

\emph{Proof:} The calculation of  $\mathbb{P}\left\{ {\tilde \gamma _{\rm{A}}^{{\rm{Ex}}}} > {\tilde \gamma _{\rm{E}}^{{\rm{Ex}}}} \right\}$ is conducted using probability theory, thus completing the proof.  \hfill $\blacksquare$

%With knowledge from probability theory, we find that $\mathbb{P}\left\{ {\tilde \gamma _{\rm{A}}^{{\rm{Ex}}}} > {\tilde \gamma _{\rm{E}}^{{\rm{Ex}}}} \right\}$ is calculated as
%\begin{align}\nonumber
%\mathbb{P}\left\{ {{\tilde \gamma _{\rm{A}}^{{\rm{Ex}}}} > {{\tilde \gamma _{\rm{E}}^{{\rm{Ex}}}}}} \right\} &= 1 - \mathbb{P}\left\{ {{{\tilde \gamma _{\rm{A}}^{{\rm{Ex}}}}} \le {{\tilde \gamma _{\rm{E}}^{{\rm{Ex}}}}}} \right\} \\\nonumber
%&= 1 - \int_0^\infty  {\int_0^{{\bar y}} {{f_{{{\tilde \gamma _{\rm{A}}^{{\rm{Ex}}}}}}}\left( {\bar x} \right){f_{{{\tilde \gamma _{\rm{E}}^{{\rm{Ex}}}}}}}\left( \bar y \right)d\bar xd\bar y} } \\
%&= 1 - \int_0^\infty  {{F_{{{\tilde \gamma _{\rm{A}}^{{\rm{Ex}}}}}}}\left( \bar y \right){f_{{{\tilde \gamma _{\rm{E}}^{{\rm{Ex}}}}}}}\left( \bar y \right)d\bar y}. \label{eq31}
%\end{align}

%Substituting (16) and \eqref{eq20} into \eqref{eq31} and then applying Gaussian-Chebyshev quadrature to prove the end.  

\textbf{\emph{Remark 3:}} As $m \to \infty $, the result in \eqref{eq28} of Theorem 1 converges to \eqref{eq30}, demonstrating the accuracy of the finite blocklength approximation proposed therein. This corollary bridges finite blocklength analysis with classical infinite blocklength information theory and quantifies the theoretical performance loss of short-packet communications relative to conventional long-packet transmission, which is dictated by parameters such as blocklength $m$ and the information leakage probability $\delta$.

\subsection{AST of AAV in Internal Eavesdropping Scenario}
Unlike the case discussed in Sec. II-B, in internal eavesdropping scenario the decoding of the AAV consists of two parts: decoding $x_{\rm{E}}$ and then decoding $x_{\rm{A}}$. Decoding errors at both stages need to be taken into account.

Thus, the BLER of the AAV is given by
\begin{align}
{\varepsilon ^{\rm{In}}} = {\varepsilon _{{\rm{A,E}}}^{{\rm{In}}}} + \left( {1 - {\varepsilon _{{\rm{A,E}}}^{{\rm{In}}}}} \right){\varepsilon _{\rm{A}}^{{\rm{In}}}} \label{eq32}
\end{align}
with
\begin{align}
{\varepsilon _{{\rm{A,E}}}^{{\rm{In}}}} &= Q\left( {\sqrt {\frac{m}{{V_{{\rm{A}} \to {\rm{E}}}^{{\rm{In}}}}}} \left( {\log \left( {1 + \tilde \gamma _{{\rm{A}} \to {\rm{E}}}^{{\rm{In}}}} \right) - \frac{B}{m}} \right)} \right) \label{eq33} \\
\varepsilon _{\rm{A}}^{{\rm{In}}} &= Q\left( {\sqrt {\frac{m}{{V_{\rm{A}}^{{\rm{In}}}}}} \left( {\log \frac{{1 + \tilde \gamma _{\rm{A}}^{{\rm{In}}}}}{{1 + \tilde \gamma _{{\rm{E}} \to {\rm{A}}}^{{\rm{In}}}}} - \iota _1 } \right)} \right) \label{eq34}
\end{align}
where ${\varepsilon _{{\rm{A,E}}}^{{\rm{In}}}}$ denotes the error probability in decoding $x_{\rm{E}}$ by the AAV; $\iota _1 = {\sqrt {{m^{ - 1}}V_{{\rm{E}} \to {\rm{A}}}^{{\rm{In}}}} {Q^{ - 1}}\left( \delta  \right)} + {m^{ - 1}}{B}$ and $\left( {1 - {\varepsilon _{{\rm{A,E}}}^{{\rm{In}}}}} \right){\varepsilon _{\rm{A}}^{{\rm{In}}}}$ denotes the probability that $x_{\rm{E}}$ is successfully decoded, but $x_{\rm{A}}$ fails to be decoded.

In practical communications, ${\varepsilon _{{\rm{A,E}}}^{{\rm{In}}}}$ is typically small and much less than one \cite{Yuehua2018wcl}. Therefore, ${\varepsilon ^{\rm{In}}}$ is approximated as
\begin{align}
{\varepsilon ^{\rm{In}}} \approx {\varepsilon _{{\rm{A,E}}}^{{\rm{In}}}} + {\varepsilon _{\rm{A}}^{{\rm{In}}}}. \label{eq35}
\end{align}

Thus, the AST of the AAV in the internal eavesdropping scenario is given by
\begin{align}
{\tau ^{\rm{In}}} = \frac{B}{m}\left( {1 - {{\bar \varepsilon }^{\rm{In}}}} \right) \label{eq36}
\end{align}
where
\begin{align}
{\bar \varepsilon ^{\rm{In}}} = \mathbb{E}\left[ {{\varepsilon _{{\rm{A,E}}}^{{\rm{In}}}}} \right] + \mathbb{E}\left[ {\varepsilon _{\rm{A}}^{{\rm{In}}}} \right]. \label{eq37}
\end{align}

Based on the above analysis of AST, we proceed to solve for $\mathbb{E}\left[ {{\varepsilon _{{\rm{A,E}}}^{{\rm{In}}}}} \right]$ and $\mathbb{E}\left[ {{\varepsilon _{\rm{A}}}} \right]$ respectively.

From the definition of expectation, we have
\begin{align}
\mathbb{E}\left[ {{\varepsilon _{{\rm{A,E}}}^{{\rm{In}}}}} \right] &= \int_0^\infty  {Q\left( {\frac{{{{\log }}\left( {1 + \bar z} \right) - \frac{B}{m}}}{{\sqrt {\frac{{V\left( {\bar z} \right)}}{m}} }}} \right){f_{\tilde \gamma _{{\rm{A}} \to {\rm{E}}}^{\rm{In}}}}\left( {\bar z} \right)d\bar z} \label{eq38}\\
\mathbb{E}\left[ {\varepsilon _{\rm{A}}^{{\rm{In}}}} \right] &= \int_0^\infty  {\int_{\hat y}^\infty {\varepsilon _{\rm{A}} \left( {\tilde z,\hat y} \right)} } {f_{\tilde z}}\left( {\tilde z} \right){f_{\hat y}}\left( {\hat y} \right)d\tilde zd\hat y \label{eq39}
\end{align}
where
\begin{align}\nonumber
\varepsilon _{\rm{A}} \left( {\tilde z,\hat y} \right) = Q&\left( {\sqrt {\frac{m}{{V_{\rm{A}}^{\rm{In}}\left( {\tilde z} \right)}}} \left( {\log \frac{{1 + \tilde z}}{{1 + \hat y}}} \right.} \right.\\
& \left. {\left. { - \sqrt {{m^{ - 1}}V_{{\rm{E}} \to {\rm{A}}}^{{\rm{In}}}\left( {\hat y} \right)}  {Q^{ - 1}}\left( \delta  \right) - \frac{B}{m}} \right)} \right). \label{eq40}
\end{align}

Note that Q-function in \eqref{eq38} making the computation of the integral intractable. Based on \eqref{eq24}, a linear approximation of the Q-function is given by \eqref{eq41} at top of next page, where $\tilde \delta  = {\big( {2\pi \big( {{2^{\frac{B}{m}}} - 1} \big)} \big)^{ - \frac{1}{2}}}$, $\tilde \beta  = {2^{\frac{B}{m}}} - 1$, $\tilde v = \tilde \beta  - {\big( {2\tilde \delta \sqrt m } \big)^{ - 1}}$ and $\tilde u = \tilde \beta  + {\big( {2\tilde \delta \sqrt m } \big)^{ - 1}}$.
\begin{figure*}[!t]
\begin{align}
\label{eq41}
Q\left( {\sqrt {\frac{m}{{V\left( {\bar z} \right)}}} \left( {{{\log }}\left( {1 + \bar z} \right) - \frac{B}{m}} \right)} \right) \approx \Phi \left( {\bar z} \right) = \left\{ \!\!\!\! {\begin{array}{*{20}{c}}
\begin{array}{l}
1,\\
0.5 + \tilde \delta \sqrt m \left( {\bar z - \tilde \beta } \right),\\
0,
\end{array}&\begin{array}{l}
\bar z \le \tilde v\\
\bar z \in \left( {\tilde v,\tilde u} \right)\\
\bar z \ge \tilde u
\end{array}
\end{array}} \right..
\end{align}
\hrulefill \vspace*{0pt}
\end{figure*}

\emph{Theorem 2:} In the internal eavesdropping scenario, the AST of the AAV is given by
\begin{align}
{\tau ^{\rm{In}}} = \frac{B}{m}\left( {1 - \mathbb{E}\left[ {{\varepsilon _{{\rm{A,E}}}^{{\rm{In}}}}} \right] - \mathbb{E}\left[\varepsilon _{\rm{A}}^{{\rm{In}}} \right]} \right) \label{eq42}
\end{align}
where
\begin{align}
\mathbb{E}\left[ {{\varepsilon _{{\rm{A,E}}}^{{\rm{In}}}}} \right]  \approx {\Gamma ^{ - 1}}\big( {\hat k} \big)\gamma \bigg( {\hat k,{{\hat \theta }^{ - 1}}\sqrt {\frac{{\tilde \beta }}{{{a_{\rm{E}}}\Delta  - {a_{\rm{A}}}\Delta \tilde \beta }}} } \bigg) \label{eq43}
\end{align}
and $\mathbb{E}\left[\varepsilon _{\rm{A}}^{{\rm{In}}} \right]$ is shown in \eqref{eq44} at the top of the next page.
In \eqref{eq44}, ${\zeta _j} = \frac{{{M_1}\left( {{\varpi _j} + 1} \right)}}{2}$, ${\varpi _j} = \cos  {\frac{{\left( {2j - 1} \right)\pi }}{{2{M_2}}}}$ and ${t_6} = {2^{\sqrt {{m^{ - 1}}\left( {1 - {{\left( {1 + {\zeta _j}} \right)}^{ - 2}}} \right)} {Q^{ - 1}}\left( \delta  \right) + \frac{B}{m}}}\left( {1 + {\zeta _j}} \right) - 1$.
\begin{figure*}[!t]
\begin{align}
\label{eq44}
\begin{aligned}
\mathbb{E}\left[\varepsilon _{\rm{A}}^{{\rm{In}}} \right] \approx \frac{{{\pi ^2}M_1^2d_{{\rm{GE}}}^\alpha }}{{4\Gamma (\hat k)\omega {\rho _{\rm{A}}}{a_{\rm{A}}}{\rho _{\rm{E}}}M_2^2}}\sum\limits_{i = 0}^{{M_2}} {{{{e}}^{ - \frac{{{\zeta _i}}}{{\omega {\rho _{\rm{A}}}}}}}\sqrt {1 - {\varpi _i}^2} } \sum\limits_{j = 0}^{{M_2}} {\gamma \left( {\hat k,{{\hat \theta }^{ - 1}}\sqrt {A_1^{ - 1}{t_6}\left( {{\zeta _i} + 1} \right)} } \right){{{e}}^{ - \frac{{{\zeta _j}}}{{{a_{\rm{A}}}{\rho _{\rm{E}}}d_{{\rm{GE}}}^{ - \alpha }}}}}\sqrt {1 - {\varpi _j}^2} } .
\end{aligned}
\end{align}
\hrulefill \vspace*{0pt}
\end{figure*} 

\emph{Proof:} The proof is provided in Appendix E.  \hfill $\blacksquare$  

\textbf{\emph{Remark 4:}} In the internal eavesdropping scenario, the residual interference from imperfect SIC introduces additional randomness, providing extra diversity gain. Consequently, the BLER with secrecy requirement in the internal eavesdropping scenario experiences more significant attenuation in the high-SNR regime compared to the external case.

Theorem 2 reveals that the analytic formulation of AST in the case of internal eavesdropping is so complicated that it is difficult to discern directly the impact of important parameters on the performance of the network. Thus, we give some asymptotic analysis by the following corollaries. The high-SNR approximation for AST at perfect SIC and imperfect SIC are given, respectively.

\emph{Corollary 3:}
When the transmit power $P_{\rm{G}}$ is sufficiently large, the AST in the high-SNR regime with perfect SIC can be expressed as follows
\begin{align}
\label{eq45}
{\tau ^{\rm{p,asy}}} \approx \frac{B}{m}\left( {1 - \mathbb{E}\big[ {\varepsilon _{{\rm{A,E}}}^{{\rm{In,asy}}}} \big] - \mathbb{E}\left[ {\varepsilon _{\rm{A}}^{\rm{p,asy}}} \right]} \right)
\end{align}
where
\begin{align}
\mathbb{E}\big[ {\varepsilon _{{\rm{A,E}}}^{{\rm{In,asy}}}} \big] &= Q\bigg( {\sqrt m \left( {\log \left( {1 + \frac{{{a_{\rm{E}}}}}{{{a_{\rm{A}}}}}} \right) - \frac{B}{m}} \right)} \bigg) \label{eq46} \\
\mathbb{E}\left[ {\varepsilon _{\rm{A}}^{\rm{p,asy}}} \right] &= {{\left( {\frac{{2{e^{{\Lambda _2}}}}}{{{\Lambda _1}{{\hat \theta }^2}}}} \right)}^{\frac{{\hat k}}{2}}}{e^{\frac{{{e^{{\Lambda _2}}}}}{{8{{\hat \theta }^2}}}}}{D_{ - \hat k}}\left( {\sqrt {\frac{{{e^{{\Lambda _2}}}}}{{2{\Lambda _1}{{\hat \theta }^2}}}} } \right). \label{eq47}
\end{align}

\emph{Proof: }In the high-SNR regime, the secrecy capacity of AAV is $\log \big( {1 + \frac{{{a_{\rm{E}}}}}{{{a_{\rm{A}}}}}} \big)$, which is a constant. Therefore, the instantaneous BLER is also a constant as shown in \eqref{eq46}. The approximation of the instantaneous BLER with secrecy requirement with perfect SIC is the same as \eqref{eqd1}. The calculation for $\mathbb{E}\left[ {\varepsilon _{\rm{A}}^{\rm{p,asy}}} \right]$ is similar.  \hfill $\blacksquare$

\emph{Corollary 4:}
When the $P_{\rm{G}}$ is sufficiently large, the AST in the high-SNR regime with imperfect SIC is given by
\begin{align}
\label{eq48}
{\tau ^{\rm{ip,asy}}} \approx 0.
\end{align}

\emph{Proof}: When the transmit power $P_{\rm{G}}$ approaches infinity, $\varepsilon _{\rm{A}}^{\rm{ip,asy}} \approx 1$. Substituting the result into \eqref{eq32}, we have ${\varepsilon ^{\rm{In,asy}}} = 1$. The proof is complete.  \hfill $\blacksquare$

\textbf{\emph{Remark 5:}} The AST with perfect SIC converges to a constant in the high-SNR regime. However, the AST with imperfect SIC actually decreases with increasing transmit power, since the residual interference power scales proportionally with the transmit power.

In the internal eavesdropping scenario, as the blocklength $m$ tends to infinity, the first stage decoding is assumed to be error-free. The second stage, however, need to account for the secrecy outage probability. The following corollary presents the closed-form expression for the secrecy throughput in infinite blocklength.

\emph{Corollary 5:} When  $m \to \infty $, the secrecy throughput is
\begin{align}\nonumber
\tau _{m \to \infty }^{{\rm{In}}}& = \frac{B}{m}\mathbb{P}\left\{ {\tilde \gamma _{\rm{A}}^{\rm{In}} > \tilde \gamma _{{\rm{E}} \to {\rm{A}}}^{\rm{In}}} \right\} \\
& \approx  \frac{B}{m}\left( {1 - {P_{\rm{A}}}} \right) \label{eq49}
\end{align}
where
\begin{align}\nonumber
\label{eq50}
{P_{\rm{A}}} =& \frac{{{\pi ^2}M_1^2d_{{\rm{GE}}}^\alpha }}{{4\Gamma (\hat k)\omega {\rho _{\rm{A}}}{a_{\rm{A}}}{\rho _{\rm{E}}}M_2^2}} \sum\limits_{i = 0}^{{M_2}} {{{{e}}^{ - \frac{{{\zeta _i}}}{{\omega {\rho _{\rm{A}}}}}}}\sqrt {1 - {\varpi _i}^2} }\\
   &\times \!\! \sum\limits_{j = 0}^{{M_2}} {\gamma \left( {\hat k,{{\hat \theta }^{ - 1}}\sqrt {A_1^{ - 1}{\zeta _j}\left( {{\zeta _i} + 1} \right)} } \right)\frac{{\sqrt {1 - {\varpi _j}^2} }}{{{e^{\frac{{{\zeta _j}}}{{{a_{\rm{A}}}{\rho _{\rm{E}}}d_{{\rm{GE}}}^{ - \alpha }}}}}}} }.
\end{align}

\textbf{\emph{Remark 6:}} As the blocklength $m \to \infty $,  the approximation of the AST \eqref{eq44} for finite blocklength detailed in Theorem 2 converges to \eqref{eq50}, demonstrating the validity and accuracy of the analytical framework established in Theorem 2.

\section{Blocklength Optimization Strategy}
Based on the analytical expressions of AST derived in Section IV, this section develops optimization strategies to maximize the AST by tuning the blocklength $m$. We first address the unconstrained optimization problem, followed by a constrained optimization framework that incorporates practical latency and reliability requirements.

\subsection{Unconstrained AST Optimization}
As mentioned in the previous section, the blocklength $m$ has an important effect on the secrecy of the AAV. In this subsection, we focus on the challenge of determining the optimal choice of blocklength when the number of transmission bits is fixed. The following theorem describes the optimal blocklength that maximizes the AST.

\emph{Theorem 3: }\label{theorem3}The AST in \eqref{eq28} is a quasi-concave function of the relaxed continuous blocklength $m$. Let ${m^*}$ denote the unique solution to the equation $\frac{{\partial {\tau ^{\rm{Ex}}}}}{{\partial m}} = 0$.  The optimal integer blocklength that maximizes the AST is obtained by evaluating the throughput at $\left\lfloor {{m^*}} \right\rfloor $ and $\left\lceil {{m^*}} \right\rceil $ and selecting the value that yields the maximum throughput.

\begin{align}
\frac{{\partial {\tau ^{\rm{Ex}}}}}{{\partial m}} \!=\! \bigg( {\frac{{B{t_4}}}{{{m^2}}}{\left( {\frac{{\Xi  - 1 + \Xi {\zeta _i}}}{{\Delta {{\hat \theta }^2}}}} \right)^{\frac{{\hat k + j}}{2}}} \!+\! \frac{{B{t_4}{t_5}}}{m}} \bigg) \!-\! \frac{B}{{{m^2}}}
\end{align}
where
\begin{align}
{t_4} &= \Lambda \sum\limits_{i = 0}^{{M_2}} {\sum\limits_{j = 0}^\infty  {\frac{{\sqrt {1 - {\varpi _i}^2} {{\left( { - 1} \right)}^j}}}{{{{{e}}^{\frac{{{\zeta _i}}}{{{\rho _{\rm{E}}}{d_{{\rm{GE}}}^{ - \alpha }}}}}}j!\big( {\hat k + j} \big)}}} } \\
{t_5} &= \frac{{\big( {\hat k + j} \big)\left( {1 + {\zeta _i}} \right){\left( {\Xi  - 1 + \Xi {\zeta _i}} \right)^{\frac{{\hat k + j - 2}}{2}}}}}{{{{\left( {\frac{{{Q^{ - 1}}\left( \delta  \right)}}{{2{m^{\frac{3}{2}}}}} + \frac{B}{{{m^2}}}} \right)}^{ - 1}}{2^{1 - \Xi }}{\big( {\Delta  {{\hat \theta }^2}} \big)^{\frac{{\hat k + j}}{2}}}\log e}}. 
\end{align}

\emph{Proof: }
First relax the natural number $m$ to a positive real number. It is easy to conclude that $\frac{{\partial {\tau ^{\rm{Ex}}}}}{{\partial m}}\left( m \right)$ decreases as $m$ increases. Moreover, easily derived ${\lim _{m \to {0^ + }}}\frac{{\partial {\tau ^{{\rm{Ex}}}}}}{{\partial m}}\left( m \right) > 0$ while ${\lim _{m \to {0^ + }}}\frac{{\partial {\tau ^{{\rm{Ex}}}}}}{{\partial m}}\left( m \right) < 0$. Therefore, when $m$ is a positive real number, the AST tends to increase and then decrease as $m$ increases. In other words, the AST is a quasi-concave function of continuous $m$. The optimal ${m^*}$ that maximizes the AST after being relaxed can be obtained by the bisection search. The proof is complete. \hfill $\blacksquare$ 

\subsection{Optimization In Reliability and Latency Constraints}
In this subsection, focusing on practical constraints, we optimize the constrained blocklength to maximize the AST, achieving a balance between latency, reliability, and secrecy.

Specifically, the optimization problem is modeled as
\begin{subequations}
\begin{align}
&\mathop {{\rm{max}}}\limits_{m \in {\mathbb{Z}^ + }} {\rm{  }}\ \ \tau ^{\rm{Ex}}  \\
&\ {\rm{s.t.}}\quad \ \bar \varepsilon \left( m \right)  \le {\varepsilon _{\rm{th}}} \tag{54b} \label{eq54b}\\
&\quad \quad \quad m \le {m_{\rm{th}}} \tag{54c} \label{eq54c}
\end{align}
\end{subequations}
where ${\varepsilon _{\rm{th}}}$ denotes the maximum BLER that can be tolerated and \eqref{eq54b} denotes the reliability constraints. \eqref{eq54c} denotes the blocklength constraint as a way to capture the constraints on the latency.

It is evident that $\bar \varepsilon $ is a decreasing function of $m$, and from Theorem 1, \eqref{eq54b} can be reformulated as a constraint in terms of $m$, that is ${\bar \varepsilon ^{ - 1}}\left( {{\varepsilon _{\rm{th}}}} \right) \le m$. Therefore, the optimization problem (54) is rewritten as
\begin{subequations}
\begin{align}
\label{eq55}
&\mathop {{\rm{max}}}\limits_{m \in {\mathbb{Z}^ + }} {\rm{  }}\ \ \tau ^{\rm{Ex}} \\
&\ {\rm{s.t.}}\quad \ \ m \ge {\bar \varepsilon ^{ - 1}}\left( {{\varepsilon _{\rm{th}}}} \right)\\
&\quad \quad \quad \  m \le {m_{\rm{th}}} \tag{55c} \label{eq55c}
\end{align}
\end{subequations}
where ${{\bar \varepsilon }^{ - 1}}\left( {{\varepsilon _{\rm{th}}}} \right)$ is the inverse function at $\bar \varepsilon \left( m \right) = {\varepsilon _{\rm{th}}}$.

Problem (55) has a solution only if $\left\lceil {{{\bar \varepsilon }^{ - 1}}\left( {{\varepsilon _{\rm{th}}}} \right)} \right\rceil $ is less than or equal to $\left\lfloor {{m_{\rm{th}}}} \right\rfloor $. The subsequent corollary presents the optimal blocklength for the optimization problem (55).

\emph{Corollary 6:} The optimal blocklength of problem (55) can be reformulated as
\begin{align}
{{\tilde m}^*} = \left\{ {\!\!\!\begin{array}{*{20}{l}}
{\left\lceil {{{\bar \varepsilon }^{ - 1}}\left( {{\varepsilon _{\rm{th}}}} \right)} \right\rceil ,}&{{m^*} \le \left\lceil {{{\bar \varepsilon }^{ - 1}}\left( {{\varepsilon _{\rm{th}}}} \right)} \right\rceil }\\
{\arg {\rm{ }}\mathop {{\rm{max}}}\limits_{{t_6}}  {\tau ^{\rm{Ex}}}\left( m \right),}&m \in \left( {\left\lceil {{{\bar \varepsilon }^{ - 1}}\left( {{\varepsilon _{{\rm{th}}}}} \right)} \right\rceil ,\left\lfloor {{m_{{\rm{th}}}}} \right\rfloor } \right)\\
{\left\lfloor {{m_{\rm{th}}}} \right\rfloor ,}&{{m^*} \ge \left\lfloor {{m_{\rm{th}}}} \right\rfloor }
\end{array}} \right.
\end{align}
where ${t_6}\! =\! m\!\! \in \left\{ \!{\left\lceil {{m^*}} \right\rceil \!\!,\left\lfloor {{m^*}} \right\rfloor }\! \right\}$ and ${{m}\!^*}$ is defined in Theorem 3.

\emph{Proof: }Theorem \ref{theorem3} demonstrates that the AST is a quasi-concave function of the relaxed continuous blocklength $m$, with a unique maximum point ${m^*}$ (obtained by solving $\frac{{\partial {\tau ^{\rm{Ex}}}}}{{\partial m}} \left( m \right) = 0$). Three cases are discussed depending on the position of ${m^*}$ with respect to the feasible interval $\left[ {\left\lceil {{{\bar \varepsilon }^{ - 1}}\left( {{\varepsilon _{\rm{th}}}} \right)} \right\rceil , \left\lfloor {{m_{\rm{th}}}} \right\rfloor } \right]$. The proof is complete. \hfill $\blacksquare$ 

\section{Numerical Results}
In this section, we present simulation results to demonstrate the secure performance of short-packet communications in AAV-RIS networks, with a focus on the secrecy performance in both external and internal eavesdropping scenarios. Unless otherwise stated, the parameters used in the simulations are as follows: ${K_{\rm{GR}}}$= 2 dB, ${K_{\rm{RA}}}$ = 2 dB, $\alpha  = 2$, ${\Omega _{\rm{GR}}} = 1$, ${\Omega _{\rm{RA}}} = 1$, ${d_{\rm{GR}}} = 30$ m, ${d_{\rm{RA}}} = 20$ m, ${d_{\rm{GE}}} = 15$ m, $N = 100$, $\delta  = {10^{-3}}$ and ${P_{\rm{G}}} = 30$ dBw. All simulations are obtained by averaging over ${10^5}$ channel realizations.
\subsection{External Eavesdropping Scenario}
In this subsection, the AST behaviors of AAV in external eavesdropping scenario are discussed.
\begin{figure}[!t]
\setlength{\abovecaptionskip}{0pt}
\centering
\includegraphics [width=3.2in]{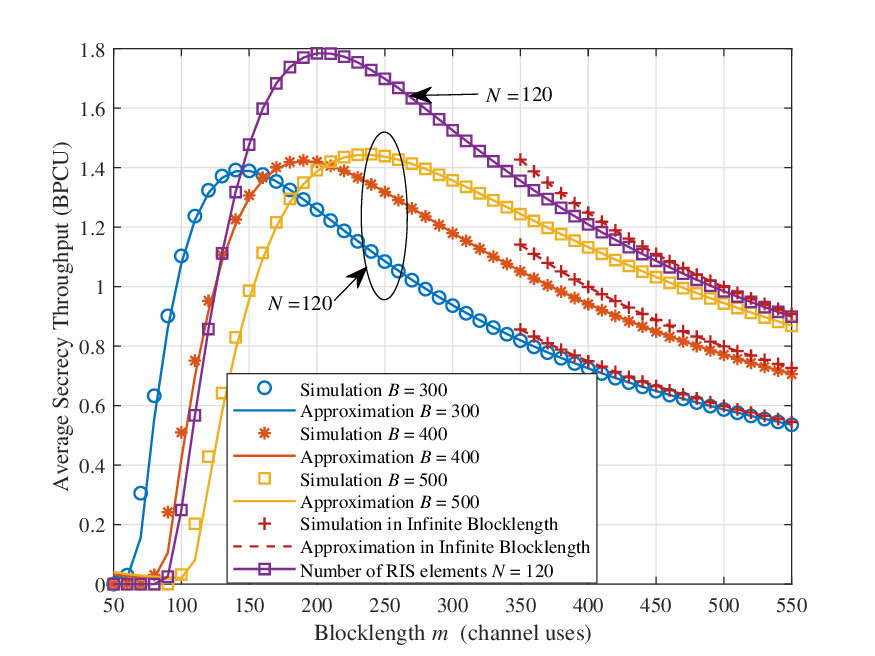}
\caption{The AST $\tau ^{\rm{Ex}}$ and its approximation in finite blocklength in the external eavesdropping scenario versus blocklength $m$ with different the number of bits transmitted per block $B$ and number of RIS elements $N$.}
\label{fig2}
\end{figure}
\begin{figure}[!t]
	\setlength{\abovecaptionskip}{0pt}
	\centering
	\includegraphics [width=3.1in]{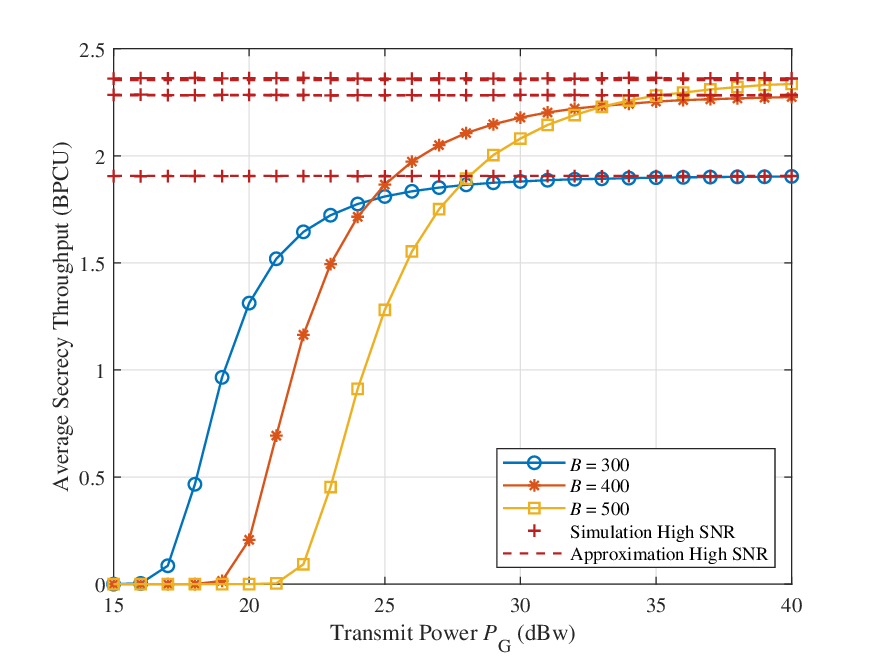}
	\caption{The AST $\tau ^{\rm{Ex}}$ and its approximation in the high-SNR regime versus the transmit power $P_{\rm{G}}$ with different the number of bits transmitted per block $B$ in the external eavesdropping scenario.}
	\label{fig3}
\end{figure}
\begin{figure}[t]
	\setlength{\abovecaptionskip}{0pt}
	\centering
	\includegraphics [width=3.2in]{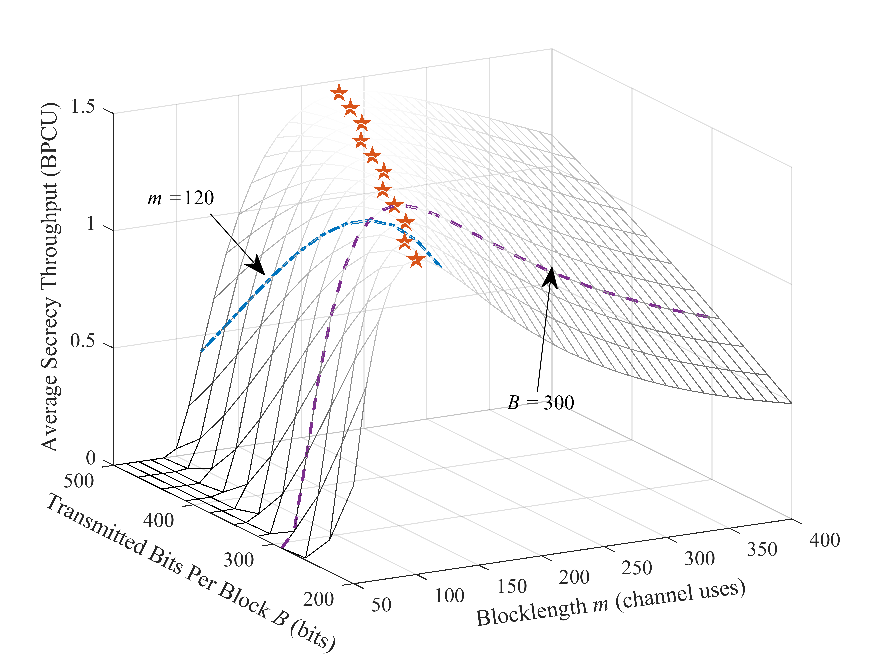}
	\caption{The AST $\tau ^{\rm{Ex}}$ versus blocklength $m$ and the number of bits transmitted per block $B$ in the external eavesdropping scenario. The pentagram markers denote the maximum AST and the corresponding blocklength ${m^*}$ in fixed the number of bits transmitted per block $B$.}
	\label{fig4}
\end{figure}
\begin{figure}[t]
	\setlength{\abovecaptionskip}{0pt}
	\centering
	\includegraphics [width=3.2in]{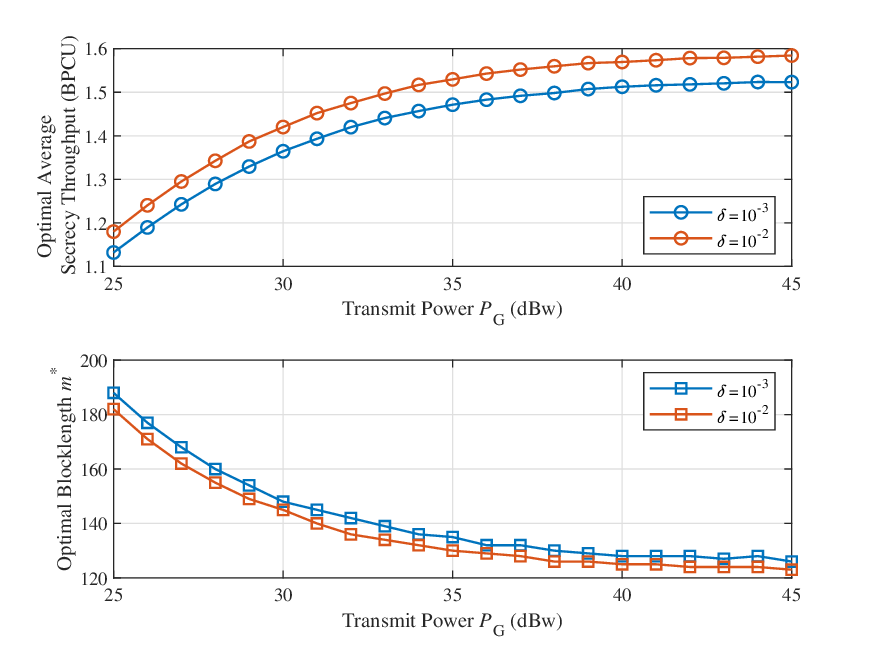}
	\caption{The optimal AST and the corresponding optimal blocklength ${m^*}$ obtained by Theorem 3 versus transmit power $P_{\rm{G}}$ with different information leakage probability $\delta$.}
	\label{fig5}
\end{figure}
The AST derived from Theorem 1 versus blocklength in the external eavesdropping scenario is illustrated in Fig. 2. The approximation curves of AST for AAV are plotted from \eqref{eq28}, which closely match the simulations. This proves that the approximations used in the integration procedure of Theorem 1 are valid. First, it can be observed that AST first increases and then decreases with blocklength $m$. When $m$ is small (left side of the peak), the BLER with secrecy requirement decreases with increasing $m$, leading to an increase in AST. When $m$ is large (right side of the peak), the BLER with secrecy requirement saturates but the spectral efficiency decreases, leading to a decrease in AST. In other words, there exists an optimal blocklength that maximizes the AST. In addition, larger the number of bits transmitted per block $B$ usually corresponds to a higher AST. The lower $B$, the shorter the optimal blocklength required, suggesting that low-load transmissions can be accomplished with shorter packets. Moreover, we present the effect of the number of RIS elements on the AST by comparing $N = $100 and $N =$ 120. The increase in the number of RIS elements has a more significant improvement in AST when the blocklength is small. When $m$ is large, the simulation shows that the AST increases only marginally with every additional 20 RIS elements, suggesting that the impact of adding more RIS elements diminishes as $m$ becomes larger. Furthermore, we compared the infinite blocklength AST and proves the correctness of Corollary 2. The infinite blocklength curve provides a theoretical upper limit for system performance, and the performance of finite blocklength curves is always lower than this limit. Moreover, the gap narrows as the $m$ increases.

The relationship between the AST and the transmit power is presented in Fig. 3. It can first be observed that the AST rises steeply with increasing transmit power $P_{\rm{G}}$ when $P_{\rm{G}}$ is low. Once $P_{\rm{G}}$ reaches a higher value (e.g., greater than 30 dBw), the growth markedly decelerates and flattens, ultimately forming a plateau. This is consistent with the result obtained in Corollary 1. In addition, it can be seen that the approximation obtained in Corollary 1 in the high-SNR regime is accurate. This indicates that the high-SNR approximation accurately predicts the true system behavior, enabling rapid assessment of ultimate performance limits while significantly reducing the computational burden of complex simulations. Moreover, we compare the impact of number of bits transmitted per block $B$ on AST. At low transmit power, smaller $B$ yields higher AST, whereas the trend reverses at high transmit power.
\begin{figure}[t]
	\setlength{\abovecaptionskip}{0pt}
	\centering
	\includegraphics [width=3.2in]{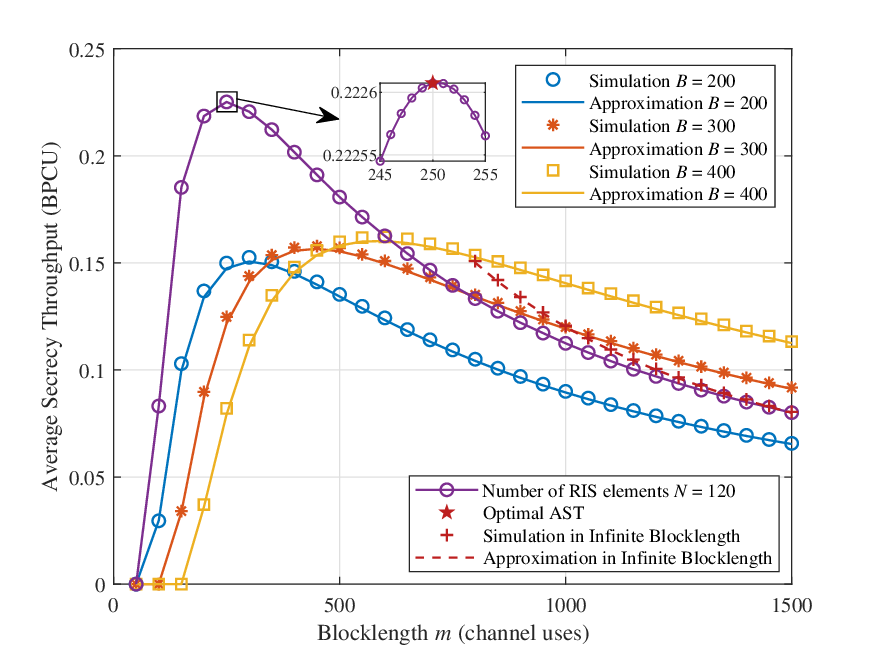}
	\caption{The AST $T$ in the internal eavesdropping scenario versus blocklength $m$ with different the number of bits transmitted per block $B$ and number of RIS elements $N$. There other system parameters are: ${a_{\rm{A}}}  =$ 0.2 and $\omega  = 0.01$.}
	\label{fig6}
\end{figure}

As shown in Fig. 4, the AST varies with both blocklength and the number of bits transmitted per block in the external eavesdropping scenario. In order to clearly show the relationship between AST with $m$ and $B$, respectively, we mark the variation of $m = 120$ and $B = 300$ with bolded curves. For the fixed $B$, the trend of AST with $m$ is consistent with Fig. 2. Similarly, For the fixed $m$, the AST first increases and then decreases with $B$. This is because the effect of $B$ on AST is twofold. On the one hand, there is a higher information transfer and on the other hand, there is a higher risk of decoding. Moreover, for each given $B$, the optimal blocklength and AST are marked with pentagram. It can be observed that the optimal blocklength ${m^*}$ increases with $B$.
\begin{figure}[!t]
	\setlength{\abovecaptionskip}{0pt}
	\centering
	\includegraphics [width=3.2in]{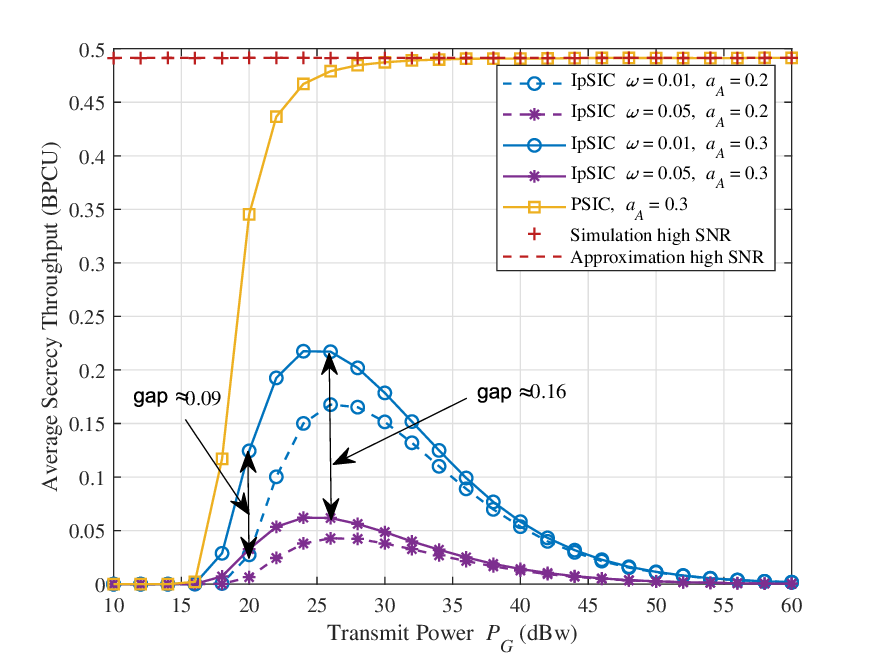}
	\caption{The AST $\tau ^{\rm{In}}$ in the internal eavesdropping scenario versus transmit power $P_{\rm{G}}$ with different residual interference level $\omega$ and power allocation coefficient $a_{\rm{A}}$. There other system parameters are: $m = 300$ and $B = 150$.}
	\label{fig7}
\end{figure}

In Fig. 5, the variations of the transmitted power $P_{\rm{G}}$ with respect to the optimal AST are presented. The first thing that can be observed that the optimal AST increases and then saturates as the transmitted power increases. This indicates that the AST is more sensitive to power in the low power range. In light of this observation, we know that a single increase in transmit power does not always lead to the desired performance improvement. In addition, we present the trend of the optimal blocklength ${m^*}$ corresponding to the optimal AST at the optimal AST with respect to the transmit power. The optimal blocklength decreases with increasing transmit power, indicating that longer blocks are needed for secrecy and reliability at low SNR. Moreover, we observe a phenomenon: a high probability of information leakage ($\delta  = {10^{ - 2}}$) corresponds to a high optimal AST. The probability of information leakage in \eqref{eq12} can be interpreted as the risk of leakage allowed by the system. Information leakage constraint relaxation allows the system to devote more resources to efficiently transmitting data, thus improving secrecy throughput. It follows from \eqref{eq12} that as $\delta$ increases and $\varepsilon$ decreases accordingly, it leads to an increase in $\tau ^{\rm{Ex}}$.

\subsection{Internal Eavesdropping Scenario}
In this subsection, the AST behaviors of AAV in internal eavesdropping scenario are discussed.

The variation between AST and blocklength in the internal eavesdropping scenario is illustrated in Fig. 6. Power allocation coefficient ${a_{\rm{A}}} = 0.2$ and residual interference level $\omega  = 0.01$ are used as examples in the simulations. The first observation is that the simulations and approximations fit well, which proves the accuracy of the approximation methods used in Theorem 2. Moreover, all ASTs first increase and then slowly decrease with $m$, which is similar to the trend in the external eavesdropping scenario. The optimal blocklength is longer for larger the number of bits transmitted per block. This is because more transmission load requires a longer blocklength to ensure performance. However, it can be observed that with the same transmission power, the AST achieved in the internal eavesdropping scenario is smaller. This is because NOMA scheme is used in this scenario, and power needs to be allocated between the two users. Furthermore, we benchmark the finite blocklength AST against its infinite blocklength counterpart. As the blocklength $m$ increases, the performance gap between the two regimes narrows. This gap is more pronounced in internal eavesdropping scenario than in external eavesdropping scenario, owing to the two-stage decoding required by the employed NOMA scheme.
\begin{figure}[!t]
	\setlength{\abovecaptionskip}{0pt}
	\centering
	\includegraphics [width=3.2in]{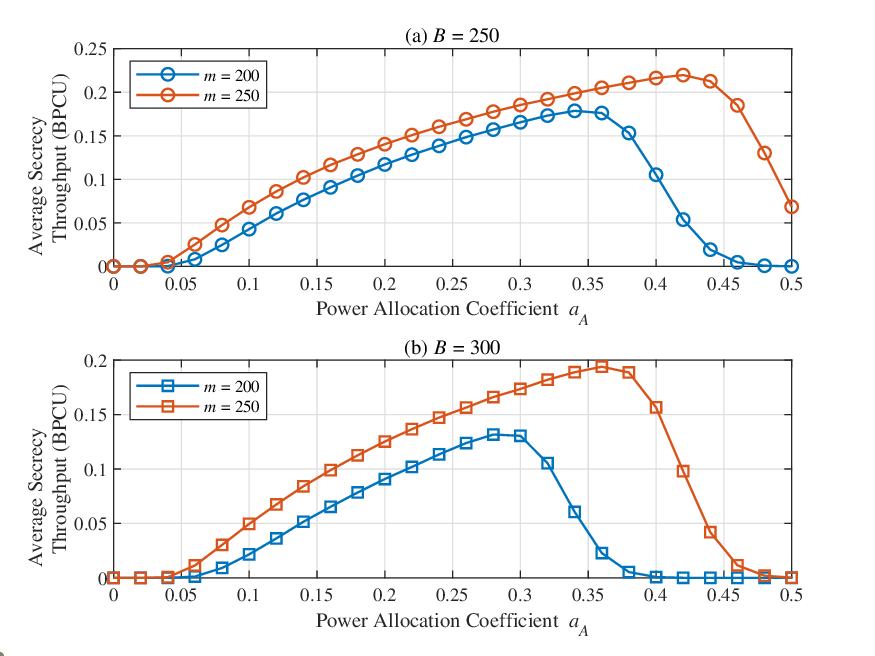}
	\caption{The AST $\tau ^{\rm{In}}$ in the internal eavesdropping scenario versus power allocation coefficient $a_{\rm{A}}$ with different the number of bits transmitted per block $B$ and blocklength $m$. There other system parameters are: transmit power ${P_{\rm{G}}} = 25$ dBw and $\omega  = 0.01$.}
	\label{fig8}
\end{figure}

The variation of AST with transmit power is illustrated in Fig. 7. In the case of imperfect SIC, the AST first increases and then decreases with the transmit power. By comparing the residual interference level $\omega  = 0.01$ and $\omega  = 0.05$ it is found that the smaller the residual interference level, the better the secrecy performance. For power allocation coefficient ${a_{\rm{A}}}  =$ 0.3, the negative impact of the residual interference level intensifies with increasing power. For instance, the performance gap is 0.09 at ${P_{\rm{G}}} = 20$ dBw, but it widens to 0.16 at ${P_{\rm{G}}} = 26$ dBw. We plot the variation of AST with transmit power in the case of perfect SIC, which tends to increase and then saturate. This suggests that SIC has a significant effect on AST. However, it should be noted that the perfect SIC has extremely high requirements for hardware and is difficult to achieve in practice. In addition, we compared the effect of different power allocation coefficient on ASTs. As expected, more power allocated to the AAV yields more AST. Moreover, we plot the asymptotic performance for the high-SNR regime. As previously noted, the AST with perfect SIC converges to a constant, whereas with imperfect SIC it becomes vanishingly small, effectively approaching zero.

The effect of power allocation coefficients on the AST is presented in Fig. 8. The $x$- axis represents the power allocation coefficient of AAV. From NOMA principles, its range is set between (0, 0.5). For any given blocklength $m$ and the number of bits transmitted per block $B$, the AST  initially increases as the power allocation coefficient $a_{\rm{A}}$ increases, reaching a peak at an optimal value. However, beyond this point, further increases in $a_{\rm{A}}$ lead to a decrease in the AST. With the overall transmission power fixed, it can be seen from \eqref{eq6} and \eqref{eq7} that the increase in the power allocation coefficient $a_{\rm{A}}$ has a dual effect. On the one hand, it makes the legitimate link more reliable, and on the other hand, it increases the SNR of the eavesdropping link. In addition, for the same power allocation coefficient $a_{\rm{A}}$, larger blocklength $m$ and the number of bits transmitted per block $B$ significantly improve the AST.

\begin{figure}[!t]
	\setlength{\abovecaptionskip}{0pt}
	\centering
	\includegraphics [width=3.2in]{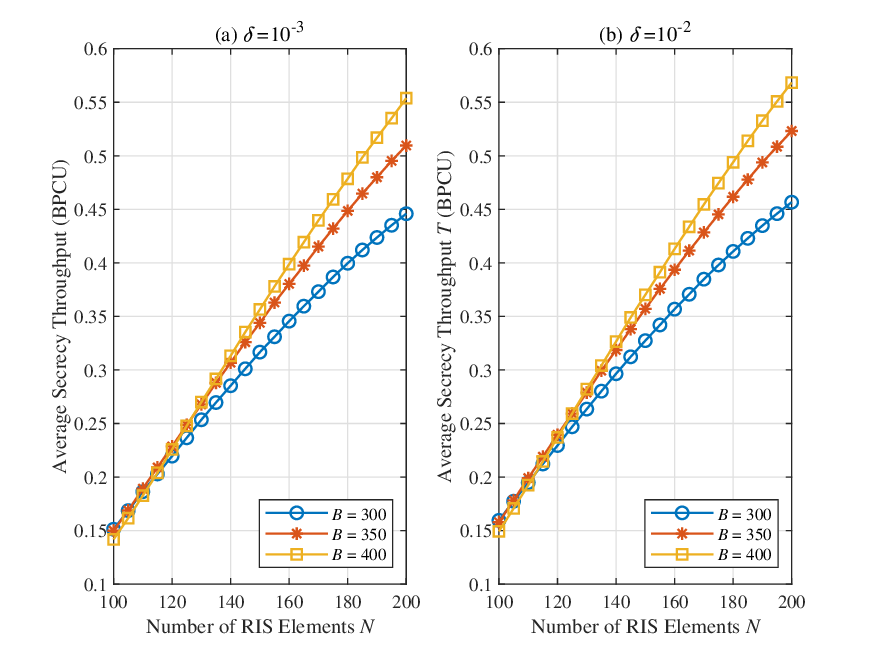}
	\caption{The AST $\tau ^{\rm{In}}$ in the internal eavesdropping scenario versus the number of RIS elements $N$ with different the number of bits transmitted per block $B$ and information leakage probability $\delta$. There other system parameters are: blocklength $m = 300$.}
	\label{fig9}
\end{figure}

The impact of the number of RIS elements on the AST is illustrated in Fig. 9. It is found that a large-scale RIS deployment positively affects the enhancement of system secrecy. This is because more RIS elements enhance the signal strength of the AAV link, resulting in a stronger and more reliable communication channel. As the signal quality increases, the system becomes more resilient to interference and eavesdropping, thereby boosting overall performance and security. Moreover, when the blocklength is fixed at $m =$ 300, the more bits transmitted per block, the larger the corresponding AST. Furthermore, by comparing the information leakage probabilities for $\delta$ = 0.1 and $\delta$ = 0.2, it has been observed that a more relaxed constraint on the leakage probability results in a higher AST. This result aligns with the trends shown in Fig. 5, where a decrease in the constraint parameter $\delta$ corresponds to an increase in the AST, indicating that loosening the information leakage limit positively impacts the performance and security of the system.

\section{Conclusion}
In this paper, we have developed comprehensive analytical frameworks for evaluating the AST of secure short-packet communications in RIS-assisted AAV networks, addressing both external and internal eavesdropping scenarios. The asymptotic behavior of AST in the high-SNR regime has been characterized, and comparisons with the infinite blocklength case have been made. Based on the derived expressions, a blocklength optimization scheme has been proposed to maximize the AST, effectively balancing the trade-offs among reliability, latency, and secrecy. The accuracy of our proposed analytical frameworks are verified by simulations, and the impact of system parameters on performance is further analyzed. The key conclusions are summarized as follows: 1) AST initially increases and then decreases with blocklength, regardless of whether it is in the external or internal eavesdropping scenario. 2) There exists an optimal blocklength that maximizes AST. 3) Increasing the transmission power does not always enhance AST as expected, especially in the internal eavesdropping scenario with imperfect SIC. 4) Large-scale RIS deployment is an effective method to improve AST.

\appendix
\section*{Appendix~A: Proof of Lemma 1}
\addcontentsline{toc}{section}{Appendix~A: Proof of Lemma 1}
\label{app:lemma1}
Let $\tilde Z = \frac{{{a_{\rm{A}}}\Delta Z}}{{\omega {{\left| {{h_{\rm{I}}}} \right|}^2}{\rho _{\rm{A}}} + 1}}$, the CDF of $\tilde \gamma _{\rm{A}}^{{\rm{In}}}$can be expressed as
\begin{align}\nonumber
\label{eqa1}
{F_{\tilde Z}}\left( {\tilde z} \right) = \int_0^\infty  {\int_0^{\frac{{\tilde z\left( {y + 1} \right)}}{{{A_1}}}} {{f_z}\left( z \right){f_y}\left( y \right)dzdy} }\\ \nonumber
  = \int_0^\infty  {{F_Z}\left( {A_1^{ - 1}\tilde z\left( {y + 1} \right)} \right){f_y}\left( y \right)dy} \tag{A.1}
\end{align}
where ${A_1} = {a_{\rm{A}}}{\rho _{\rm{A}}}{d_{{\rm{RA}}}^{ - \alpha }d_{{\rm{GR}}}^{ - \alpha }} = {a_{\rm{A}}}\Delta $. Let $\Omega _{\rm{I}} = 1$, hence the PDF of $Y = \omega {\left| {{h_{\rm{I}}}} \right|^2}{\rho _{\rm{A}}}$ is given by
\begin{align}
\label{eqa2}\nonumber
{f_Y}\left( y \right) = {\left( {\omega {\rho _{\rm{A}}}} \right)^{ - 1}}{e^{ - \frac{y}{{\omega {\rho _{\rm{A}}}}}}}. \tag{A.2}
\end{align}
Substituting \eqref{eq15} and \eqref{eqa2} into \eqref{eqa1}, we have
\begin{align} \nonumber
\label{eqa3}
{F_{\tilde Z}}\left( {\tilde z} \right) = \frac{{\int_0^\infty  {\gamma \Big( {\hat k,{{\hat \theta }^{ - 1}}\sqrt {A_1^{ - 1}\tilde z\left( {y + 1} \right)} } \Big){e^{ - \frac{y}{{\omega {\rho _{\rm{A}}}}}}}dy} }}{{\Gamma \big( {\hat k} \big)\omega {\rho _{\rm{A}}}}} . \tag{A.3}
\end{align}

It is difficult to compute \eqref{eqa3} since the presence of the lower incomplete Gamma function significantly increases the complexity of the expression. Let ${\cal L}\left( y \right)\! = \gamma \Big( {\hat k,{{\hat \theta }^{ - 1}}\sqrt {A_1^{ - 1}\tilde z\left( {y + 1} \right)}  } \Big){e^{ - \frac{y}{{\omega {\rho _{\rm{A}}}}}}}$, we have
\begin{align} \nonumber
\label{eqa4}
{\lim _{y \to \infty }}{\cal L}\left( y \right) = 0. \tag{A.4}
\end{align}
There are ${\int_{M_1}^\infty  {\gamma \Big( {\hat k,{{\hat \theta }^{ - 1}}\sqrt {A_1^{ - 1}\tilde z\left( {y + 1} \right)} } \Big){e^{ - \frac{y}{{\omega {\rho _{\rm{A}}}}}}}dy} } \to 0$ when $M_1$ is sufficiently large. Thus, the integral in \eqref{eqa3} is reformulated as
\begin{align} \nonumber
\int_0^{{M_1}} {\gamma \left( {\hat k,{{\hat \theta }^{ - 1}}\sqrt {A_1^{ - 1}\tilde z\left( {y + 1} \right)} } \right){e^{ - \frac{y}{{\omega {\rho _{\rm{A}}}}}}}dy}  . \tag{A.5}
\end{align}

Then applying Gaussian-Chebyshev quadrature \cite{handbook} and substituting the result into \eqref{eqa3}, we can get \eqref{eq19}.

\section*{Appendix~B: Proof of Lemma 2}
Denote the approximation error as
\begin{align}\nonumber
\tilde \Psi \left( \bar y \right) \approx \int_0^{\bar y}  {\big( {1 - {\varepsilon _{{{\tilde \gamma _{\rm{A}}^{{\rm{Ex}}}}} = \bar x|{{\tilde \gamma _{\rm{E}}^{{\rm{Ex}}}}}}}\left(\bar x \right)} \big){f_{{{\tilde \gamma _{\rm{A}}^{{\rm{Ex}}}}}}}\left(\bar x \right)d \bar x}.
\tag{B.1}
\end{align}
When ${{\tilde \gamma _{\rm{A}}^{{\rm{Ex}}}}} < {{\tilde \gamma _{\rm{E}}^{{\rm{Ex}}}}}$, we have $\log \frac{{1 + {{\tilde \gamma _{\rm{A}}^{{\rm{Ex}}}}}}}{{1 + {{\tilde \gamma _{\rm{E}}^{{\rm{Ex}}}}}}} < 0$ and $\sqrt {\frac{{{V_{\rm{E}}}}}{{{V_{\rm{A}}}}}}  > 1$. In addition, there are ${V_{\rm{E}}} \le 1$. 

Therefore, the lower bound of $\varepsilon $ is given by
\begin{align}\nonumber
\varepsilon  \ge Q\big( { - {Q^{ - 1}}\left( \delta  \right) - B{m^{ - \frac{1}{2}}}} \big).
\tag{B.2}
\end{align}
Then, the result of $\tilde \Psi \left( \bar y \right)$ is expressed as
\begin{align}\nonumber
\tilde \Psi \left( \bar y \right) \le \left( {1 - Q\big( { - {Q^{ - 1}}\left( \delta  \right) - B{m^{ - \frac{1}{2}}}} \big)} \right){F_{{\tilde \gamma _{\rm{A}}}}}\left( \bar y \right).
\tag{B.3}
\end{align}
Substituting some typical parameters, $m =$ 100, $B =$ 300 and $\delta  = {10^{ - 3}}$. We have $Q\left( {33.1} \right) \to 0$ and $Q\left( { - 33.1} \right) = 1 - Q\left( {33.1} \right) \to 1$. Therefore, the approximation in Lemma 2 is tight. The proof is complete.

\section*{Appendix~C: Proof of Theorem 1}
From \eqref{eq24} and the approximate discussion, $\Psi \left( \bar y \right)$ is calculated by the following
\begin{align}\nonumber
\Psi \! \left( \bar y \right)\! \approx 1 \!-& \bigg( {\underbrace {\int_0^{\frac{1}{{2k}} + {x_0}} {{f_{{{\tilde \gamma _{\rm{A}}^{{\rm{Ex}}}}}}}\left(\bar x \right)dx} }_{{S_1}}} \bigg. \\
&\ + \!\!\bigg.\! {\underbrace {\int_{\frac{1}{{2k}} + {x_0}}^{ - \frac{1}{{2k}} + {x_0}}\! {\left( {\frac{1}{2} + k\left(\! {\bar x \!- {x_0}} \!\right)} \right)\!{f_{{{\tilde \gamma _{\rm{A}}^{{\rm{Ex}}}}}}}\big(\!\bar x \!\big)d \bar x} }_{{S_2}}} \!\bigg).
\tag{C.1}
\end{align}
The result of $S_1$ is ${F_{{{\tilde \gamma _{\rm{A}}^{{\rm{Ex}}}}}}}\left( {\frac{1}{{2k}} + {x_0}} \right)$. Applying partial integration to $S_2$, we have ${S_2} =  - {F_{{{\tilde \gamma _{\rm{A}}^{{\rm{Ex}}}}}}}\left( {\frac{1}{{2k}} + {x_0}} \right) - k\int_{\frac{1}{{2k}} + {x_0}}^{ - \frac{1}{{2k}} + {x_0}} {{F_{{{\tilde \gamma _{\rm{A}}^{{\rm{Ex}}}}}}}\left(\bar x \right)d \bar x} $. Hence,
\begin{align}\nonumber
\Psi \left( \bar y \right) \approx 1 + k\int_{\frac{1}{{2k}} + {x_0}}^{ - \frac{1}{{2k}} + {x_0}} {{F_{{{\tilde \gamma _{\rm{A}}^{{\rm{Ex}}}}}}}\left(\bar x \right)d \bar x}
\tag{C.2}
\end{align}

Further, leveraging the first-order Riemann integral approximation $\int_a^b {f\left( x \right)dx}  \approx \left( {b - a} \right)f\left( {\frac{{a + b}}{2}} \right)$, we have
\begin{align}\nonumber
\Psi \left( \bar y \right){\rm{ }}\approx 1 - {F_{{{\tilde \gamma _{\rm{A}}^{{\rm{Ex}}}}}}}\left( {{x_0}} \right)
\tag{C.3} \label{eqc3}
\end{align}
Therefore, AST can be obtained by substituting \eqref{eqc3} into \eqref{eq22} as following
\begin{align}\nonumber
\tau ^{\rm{Ex}} \approx \frac{B}{m}\int_0^\infty  {\left( {1 - {F_{{{\tilde \gamma _{\rm{A}}^{{\rm{Ex}}}}}}}\left( {{x_0}\left( \bar y \right)} \right)} \right){f_{{{\tilde \gamma _{\rm{E}}^{{\rm{Ex}}}}}}}\left( \bar y \right)d{\bar y}}.
\tag{C.4} \label{eqc4}
\end{align}
When the SNR of Eve is sufficiently large, ${V_{\rm{E}}}$ approaches one and ${x_0} \approx {2^{{m^{\frac{1}{2}}}{Q^{ - 1}}\left( \delta  \right) + \frac{B}{m}}}\left( {1 + \tilde \gamma _{\rm{E}}^{{\rm{Ex}}}} \right) - 1$. Then upon substituting \eqref{eq16} and \eqref{eq25} into \eqref{eqc4}, we obtain
\begin{align}\nonumber
\tau ^{\rm{Ex}} \approx \frac{B}{m}\left( {1 - \frac{{\int_0^\infty  {\gamma \left( {\hat k,\sqrt {\frac{{\Xi  - 1 + \Xi \bar y}}{{\Delta {{\hat \theta }^2}}}} } \right){e^{ - \frac{{\bar y}}{{{\rho _{\rm{E}}}d_{{\rm{GE}}}^{ - \alpha }}}}}d\bar y} }}{{\Gamma (\hat k){\rho _{\rm{E}}}d_{{\rm{GE}}}^{ - \alpha }}}} \right).
\tag{C.5} \label{eqc5}
\end{align}
It is important to note that the form of the integral in \eqref{eqc5} is similar to that of ${\cal L}\left( y \right)$ in \eqref{eqa4}. Therefore, using the same method, the upper limit of the integral is truncated using the larger value $M_1$. The proof is complete by utilizing the Gaussian-Chebyshev quadrature \cite{handbook}.

\section*{Appendix~D: Proof of Corollary 1}
In the high-SNR regime, the instantaneous BLER with secrecy requirement approximation is
\begin{align}
{\varepsilon ^{asy}} = Q\left( {\sqrt m \left( {\log  \iota _2 - {\Lambda _2}} \right)} \right).
\tag{D.1} \label{eqd1}
\end{align}
where $\iota _2 =  {\Lambda _1}{\big( {\sum\nolimits_{i = 1}^N {\left| {{h_{{\rm{GR}},i}}{h_{{\rm{RA}},i}}} \right|} } \big)^2}{\left| {{h_{{\rm{GE}}}}} \right|^{ - 2}}$.

Let $\tilde y = {\left| {{h_{{\rm{GE}}}}} \right|^2}$. Borrowing from \eqref{eq24}, using the first-order Taylor approximation for the Q-function, with the expansion point and slope being ${z_0} = {\Lambda _1^{ - 1}{e^{{\Lambda _2}}}\tilde y}$ and $\tilde k = \frac{{\partial {\varepsilon ^{\rm{asy}}}\left( z \right)}}{{\partial z}}{|_{z = {z_0}}} =  - \sqrt {\frac{m}{{2\pi z_0^2}}} $, respectively. In a similar way to Appendix A, the average BLER with secrecy requirement can be obtained as
\begin{align}
{{\bar \varepsilon }^{{\rm{asy}}}} \approx {\Gamma ^{ - 1}}\big( {\hat k} \big)\int_0^\infty  {\gamma \bigg( {\hat k,{{\hat \theta }^{ - 1}}\sqrt {\Lambda _1^{ - 1}{e^{{\Lambda _2}}}\tilde y}  } \bigg){e^{ - \tilde y}}d\tilde y} .
\tag{D.2}
\end{align}

Utilizing \cite[(6.454)]{Table}, we are able to derive \eqref{eq29}. Thus, the proof is complete.

\section*{Appendix~E: Proof of Theorem 2}
We solve for $\mathbb{E}\left[ {{\varepsilon _{{\rm{A,E}}}^{{\rm{In}}}}} \right]$ and $\mathbb{E}\left[ {{\varepsilon _{\rm{A}}}} \right]$ respectively. From \eqref{eq38} and the approximation of the Q-function in \eqref{eq41}, the average BLER can be approximated as
\begin{align}\nonumber
 \mathbb{E}\left[ {{\varepsilon _{{\rm{A,E}}}^{{\rm{In}}}}} \right] & \approx  \tilde \delta \sqrt m \int_{\tilde v}^{\tilde u} {{F_{\tilde \gamma _{{\rm{A}} \to {\rm{E}}}^{\rm{In}}}}\left( \bar z \right)d\bar z}  \\
 &= \frac{{\tilde \delta \sqrt m }}{{\Gamma \big( {\hat k} \big)}}\int_{\tilde v}^{\tilde u} {\gamma \left( {\hat k,{{\hat \theta }^{ - 1}}\sqrt {\frac{{\bar z}}{{{a_{\rm{E}}}\Delta  - {a_{\rm{A}}}\Delta \bar z}}} } \right)d\bar z}
\tag{E.1} \label{eqe1}
\end{align}
which is obtained by using the partial integration method.

Using the first-order Riemann integral approximation for \eqref{eqe1}, we can obtain \eqref{eq43}. Then, for $\mathbb{E}\left[ {{\varepsilon _{\rm{A}}^{{\rm{In}}}}} \right]$, using the approach similar to that of Theorem 1, we have
\begin{align}\nonumber
\mathbb{E}\left[ {{\varepsilon _{\rm{A}}^{{\rm{In}}}}} \right] \approx {\int_0^\infty  {{F_{\tilde Z}}\left( {{x_0}\left(\hat y \right)} \right){f_{{\tilde \gamma _{\rm{E}}}}}\left(\hat y \right)d\hat y} }.
\tag{E.2} \label{eqe2}
\end{align}

In contrast to \eqref{eq3}, $\tilde \gamma _{{\rm{E}} \to {\rm{A}}}^{\rm{In}}$ in \eqref{eq7} is multiplied by a coefficient $a_{\rm{A}}$ less than 0.5, and thus the channel dispersion of the eavesdropping channel $V_{\rm{E}}^{\rm{In}}$ cannot be analogized to the approximation used in computing \eqref{eqe2}. By directly applying the Gaussian-Chebyshev quadrature rule along with standard mathematical manipulations, the closed-form expression in \eqref{eq44} is obtained. Finally, by substituting \eqref{eq43} and \eqref{eq44} into \eqref{eq42}, the proof is complete.

\bibliographystyle{IEEEtran}
\bibliography{lhlbib}
\end{document}